\documentclass[twocolumn]{grant_aastex}

\usepackage{verbatim}
\usepackage{morefloats}
\usepackage{apjfonts}
\usepackage{amsmath}
\usepackage{grant_defs}

\shortauthors{Tremblay et al.}
\shorttitle{A Cold Molecular Fountain in Abell 2597}

\begin{document}

\title{A GALAXY-SCALE FOUNTAIN OF COLD MOLECULAR GAS PUMPED BY A BLACK HOLE}

\vspace*{-5mm}

\author[0000-0002-5445-5401]{G.~R.~Tremblay}
\altaffiliation{Einstein Fellow}
\affiliation{Harvard-Smithsonian Center for Astrophysics, 60 Garden St., Cambridge, MA 02138, USA}
\affiliation{Yale Center for Astronomy and Astrophysics, Yale University, 52 Hillhouse Ave., New Haven, CT 06511, USA}

\author[0000-0003-2658-7893]{F.~Combes}
\affiliation{LERMA, Observatoire de Paris, PSL Research Univ.,
 College de France, CNRS, Sorbonne Univ.,  Paris, France}

\author{J.~B.~R.~Oonk}
\affiliation{ASTRON, Netherlands Institute for Radio Astronomy,
 P.O. Box 2, 7990 AA Dwingeloo, The Netherlands}
\affiliation{Leiden Observatory, Leiden University, P.O. Box 9513,
 2300 RA Leiden, The Netherlands}

\author{H.~R.~Russell}
\affiliation{Institute of Astronomy, Cambridge University, Madingley Rd., Cambridge, CB3 0HA, UK}

\author[0000-0001-5226-8349]{M.~A.~McDonald}
\affiliation{Kavli Institute for Astrophysics and Space Research,
 Massachusetts Institute of Technology, 77 Massachusetts Avenue, Cambridge, MA 02139, USA}

\author[0000-0003-2754-9258]{M.~Gaspari}
\altaffiliation{Einstein \& Spitzer Fellow}
\affiliation{Department of Astrophysical Sciences, Princeton University, Princeton, NJ
 08544, USA}

\author[0000-0003-2901-6842]{B.~Husemann}
\affiliation{Max-Planck-Institut f\"{u}r Astronomie, K\"onigstuhl 17, D-69117 Heidelberg, Germany}

\author[0000-0003-0297-4493]{P.~E.~J. Nulsen}
\affiliation{Harvard-Smithsonian Center for Astrophysics, 60 Garden St., Cambridge, MA 02138, USA}
\affiliation{ICRAR, University of Western Australia, 35 Stirling Hwy, Crawley, WA 6009, Australia}

\author[0000-0002-2622-2627]{B.~R.~McNamara}
\affiliation{Physics \& Astronomy Department, Waterloo University,
 200 University Ave.~W., Waterloo, ON, N2L, 2G1, Canada}

\author{S.~L.~Hamer}
\affiliation{CRAL, Observatoire de Lyon, CNRS, Universit\'{e} Lyon 1, 9 Avenue Ch.~Andr\'{e}, 69561 Saint-Genis-Laval, France}

\author{C.~P.~O'Dea}
\affiliation{Department of Physics \& Astronomy, University of Manitoba, Winnipeg, MB R3T 2N2, Canada}
\affiliation{School of Physics \& Astronomy, Rochester Institute of Technology,
 84 Lomb Memorial Drive, Rochester, NY 14623, USA}

\author{S.~A.~Baum}
\affiliation{School of Physics \& Astronomy, Rochester Institute of Technology,
 84 Lomb Memorial Drive, Rochester, NY 14623, USA}
\affiliation{Faculty of Science, University of Manitoba, Winnipeg, MB R3T 2N2, Canada}

\author[0000-0003-4932-9379]{T.~A.~Davis}
\affiliation{School of Physics \& Astronomy, Cardiff University,
 Queens Buildings, The Parade, Cardiff, CF24 3AA, UK}

\author[0000-0002-2808-0853]{M.~Donahue}
\author[0000-0002-3514-0383]{G.~M.~Voit}
\affiliation{Michigan State University, Physics and Astronomy Dept., East Lansing, MI 48824-2320, USA}

\author[0000-0002-3398-6916]{A.~C.~Edge}
\affiliation{Department of Physics, Durham University, Durham, DH1 3LE, UK}

\author{E.~L.~Blanton}
\affiliation{Astronomy Department and Institute for Astrophysical
 Research, Boston University, 725 Commonwealth Ave., Boston, MA 02215, USA}

\author{M.~N.~Bremer}
\affiliation{H.~W.~Wills Physics Laboratory, University of Bristol,
 Tyndall Avenue, Bristol, BS8 1TL, UK}

\author{E.~Bulbul}
\affiliation{Harvard-Smithsonian Center for Astrophysics, 60 Garden St., Cambridge, MA 02138, USA}

\author[0000-0001-6812-7938]{T.~E.~Clarke}
\affiliation{Naval Research Laboratory Remote Sensing Division, Code 7213
 4555 Overlook Ave SW, Washington, DC 20375, USA}

\author{L.~P.~David}
\affiliation{Harvard-Smithsonian Center for Astrophysics, 60 Garden St., Cambridge, MA 02138, USA}

\author{L.~O.~V.~Edwards}
\affiliation{Physics Department, California Polytechnic State University, San Luis Obispo, CA 93407, USA}

\author{D.~Eggerman}
\affiliation{Yale Center for Astronomy and Astrophysics, Yale University, 52 Hillhouse Ave., New Haven, CT 06511, USA}

\author[0000-0002-9378-4072]{A.~C.~Fabian}
\affiliation{Institute of Astronomy, Cambridge University, Madingley Rd., Cambridge, CB3 0HA, UK}

\author[0000-0002-9478-1682]{W.~Forman}
\affiliation{Harvard-Smithsonian Center for Astrophysics, 60 Garden St., Cambridge, MA 02138, USA}

\author{C.~Jones}
\affiliation{Harvard-Smithsonian Center for Astrophysics, 60 Garden St., Cambridge, MA 02138, USA}

\author{N.~Kerman}
\affiliation{Yale Center for Astronomy and Astrophysics, Yale University, 52 Hillhouse Ave., New Haven, CT 06511, USA}

\author[0000-0002-0765-0511]{R.~P.~Kraft}
\affiliation{Harvard-Smithsonian Center for Astrophysics, 60 Garden St., Cambridge, MA 02138, USA}

\author{Y.~Li}
\affiliation{Center for Computational Astrophysics, Flatiron Institute,
 162 Fifth Ave., New York, NY 10027, USA}
\affiliation{Department of Astronomy, University of Michigan, 1085 S.~University Ave.,
 Ann Arbor, MI 48109, USA}

\author[0000-0003-2284-8603]{M.~Powell}
\affiliation{Yale Center for Astronomy and Astrophysics, Yale University, 52 Hillhouse Ave., New Haven, CT 06511, USA}

\author[0000-0002-3984-4337]{S.~W.~Randall}
\affiliation{Harvard-Smithsonian Center for Astrophysics, 60 Garden St., Cambridge, MA 02138, USA}

\author{P.~Salom\'{e}}
\affiliation{LERMA, Observatoire de Paris, PSL Research Univ.,
 College de France, CNRS, Sorbonne Univ.,  Paris, France}

\author{A.~Simionescu}
\affiliation{Institute of Space and Astronautical Science (ISAS), JAXA, 3-1-1
 Yoshinodai, Chuo-ku, Sagamihara, Kanagawa, 252-5210, Japan}

\author{Y.~Su}
\affiliation{Harvard-Smithsonian Center for Astrophysics, 60 Garden St., Cambridge, MA 02138, USA}

\author[0000-0001-5880-0703]{M.~Sun}
\affiliation{Department of Physics \& Astronomy, University of Alabama in Huntsville, Huntsville, AL 35899, USA}

\author[0000-0002-0745-9792]{C.~M.~Urry}
\affiliation{Yale Center for Astronomy and Astrophysics,
 Yale University, 52 Hillhouse Ave., New Haven, CT 06511, USA}

\author[0000-0003-4227-4838]{A.~N.~Vantyghem}
\affiliation{Physics \& Astronomy Department, Waterloo University,
 200 University Ave.~W., Waterloo, ON, N2L, 2G1, Canada}

\author[0000-0003-1809-2364]{B.~J.~Wilkes}
\affiliation{Harvard-Smithsonian Center for Astrophysics, 60 Garden St., Cambridge, MA 02138, USA}

\author{J.~A.~ZuHone}
\affiliation{Harvard-Smithsonian Center for Astrophysics, 60 Garden St., Cambridge, MA 02138, USA}

\begin{abstract}
 We present ALMA and MUSE observations of the Brightest Cluster Galaxy in
 Abell 2597, a nearby ($z=0.0821$) cool core cluster of galaxies. The data map the kinematics of a three billion solar mass filamentary nebula that spans the innermost 30 kpc of the galaxy's core. Its warm ionized and cold molecular components are both cospatial and comoving, consistent with the hypothesis that the optical nebula traces the warm envelopes of many cold molecular clouds that drift in the velocity field of the hot X-ray atmosphere.
 The clouds are not in dynamical equilibrium, and instead
 show evidence for inflow toward the central supermassive black hole, outflow along the jets it launches, and uplift by the buoyant hot bubbles those jets inflate. The entire scenario is therefore consistent with a galaxy-spanning ``fountain'',
 wherein cold gas clouds drain into the black hole accretion reservoir, powering jets and bubbles that uplift a cooling plume of low-entropy multiphase gas, which may stimulate additional cooling and accretion as part of a self-regulating feedback loop. All velocities are below the escape speed from the galaxy, and so these clouds should rain back toward the galaxy center from which they came, keeping the fountain long-lived. The data are consistent with major predictions of chaotic cold accretion, precipitation, and stimulated feedback models, and may trace processes fundamental to galaxy evolution at effectively all mass scales.
\end{abstract}


\section{Introduction} \label{sec:intro}

\begin{figure*}
 \begin{center}
  \includegraphics[width=\textwidth]{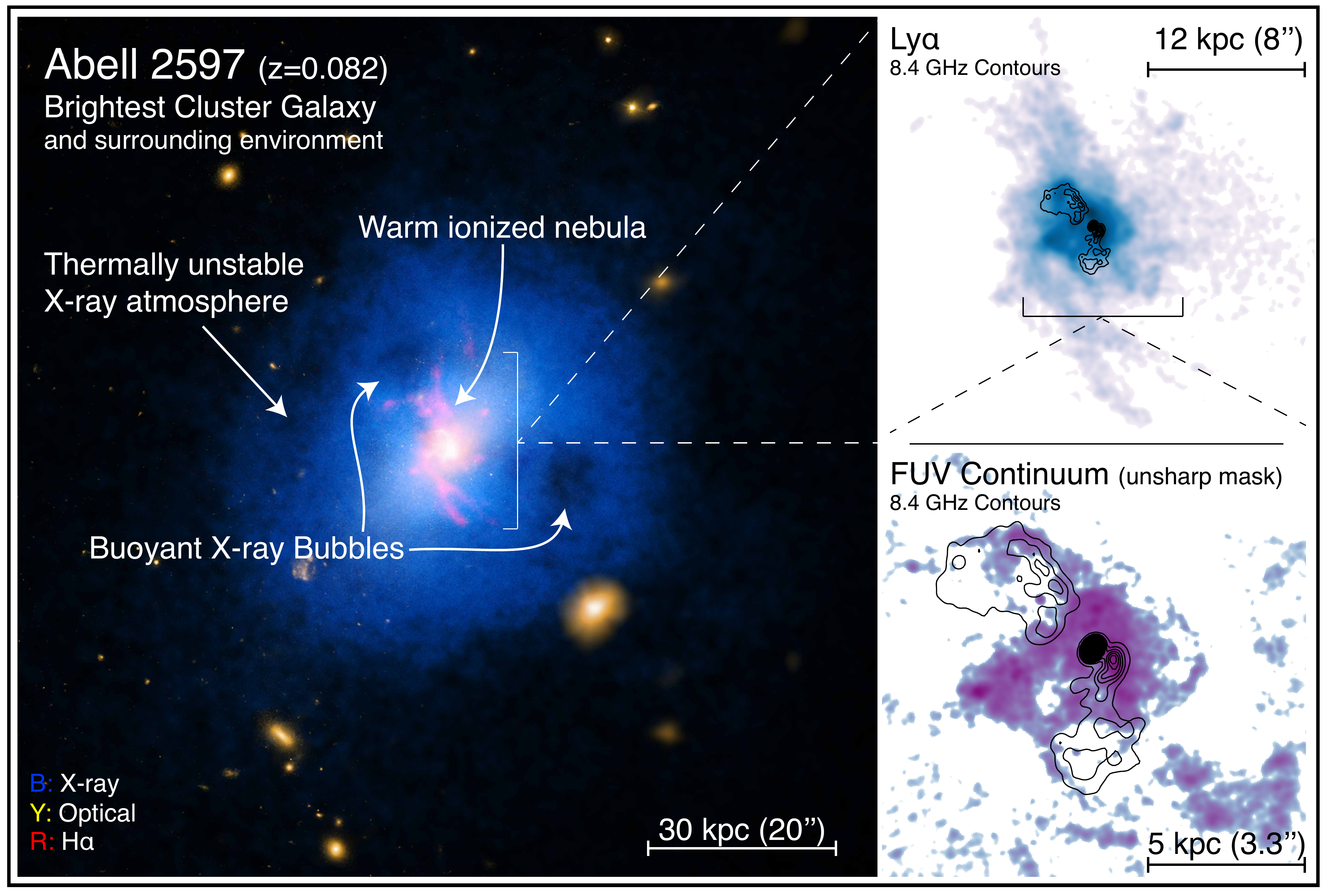}
 \end{center}
 \vspace*{-5mm}
 \caption{
 A multiwavelength view of the Abell 2597 Brightest Cluster Galaxy.  (\textit{Left})  \textit{
  Chandra} X-ray, \textit{HST} and DSS optical, and  Magellan H$\alpha$+[N~\textsc{
  ii}] emission  is shown in blue, yellow, and red, respectively {\small
   (Credit: X-ray: NASA/CXC/Michigan State Univ/G.Voit et al; Optical: NASA/STScI
   \& DSS; H$\alpha$: Carnegie Obs./Magellan/W.Baade
   Telescope/U.Maryland/M.McDonald)}.  (\textit{Top right}) \textit{HST}/STIS MAMA image of
 Ly$\alpha$ emission  associated with the ionized gas nebula. Very Large Array
 (VLA) radio contours  of the 8.4 GHz source are overlaid in black.  (\textit{Bottom right}) Unsharp mask of the \textit{HST}/ACS SBC far-ultraviolet continuum image  of the central regions of the nebula. 8.4 GHz contours are once again overlaid.
 In projection, sharp-edged rims of FUV continuum to the north and south
 wrap around the edges of the
 radio lobes.   Dashed lines indicate relative fields of view between each panel.
 The centroids  of all panels are aligned, with East left and North
 up. This figure has been partially adapted from \citealt{tremblay16}.
 }
 \label{fig:overview}
\end{figure*}

Abell 2597 is a cool core cluster of galaxies at redshift $z=0.0821$
(\autoref{fig:overview}). The galaxies inhabit a megaparsec-scale bath of X-ray
bright, $\sim 10^{7-8}$ K plasma whose central particle density is sharply
peaked about a giant elliptical brightest cluster galaxy (BCG) in the cluster
core.   Under the right conditions (e.g., \citealt{fabian94, peterson06}), the
dense halo of plasma that surrounds this galaxy can act
like a reservoir from which hot gas rapidly cools, driving a long-lived rain of thermally unstable
multiphase gas that collapses toward the galaxy's center (e.g., \citealt{gaspari17b}), powering black hole
accretion and $\sim 5$ \Msol\ yr\mone\ of star formation
\citep{tremblay12a,tremblay16}. The rate at which these cooling flow mass sinks accumulate
would likely  be higher were
the hot atmosphere not permeated by a $\sim30$ kpc-scale network of buoyantly
rising bubbles  (\autoref{fig:overview}\textit{a}),  inflated by the propagating jet
launched by the BCG's central accreting supermassive black hole
\citep{taylor99,mcnamara01,clarke05,tremblay12b}.
Those clouds that have managed to cool now form a multiphase filamentary nebula, replete with young stars, that spans the inner $\sim30$ kpc of the galaxy.
Its fractal tendrils, likely made of many cold molecular clouds with warmer ionized envelopes (e.g., \citealt{jaffe05}), wrap
around both the radio jet and the the X-ray cavities the jet has inflated
(\autoref{fig:overview}\textit{b}/\textit{c}, \citealt{mcnamara93, voit97, koekemoer99,
 mcnamara99, odea04,oonk10,tremblay12a, tremblay15, mittal15}).

These X-ray cavities act as a calorimeter for the efficient coupling between the
kinetic energy of the jet and the hot intracluster medium through which it
propagates (e.g., \citealt{churazov01,churazov02}). Given their ubiquity in effectively all cool core
clusters, systems like Abell 2597 are canonical examples of mechanical
black hole feedback, a model now routinely invoked to reconcile observations
with a theory that would otherwise over-predict  the size of galaxies and the
star formation history of the Universe (see, e.g., reviews by
\citealt{veilleux05,mcnamara07,mcnamara12,fabian12,alexander12,kormendy13,gaspari13,bykov15}). Yet, just as for quasar-driven radiative feedback invoked at earlier epochs (e.g.,
\citealt{croton06,bower06}), the degree to which the mechanical luminosity of jets
might quench (or even trigger) star formation depends on how it might couple to
the origin and fate of cold molecular gas, from which all stars are born.

Observational evidence for this coupling grows even in the absence of a
consensus explanation for it.
The density contrast between hot ($\sim10^7$ K) plasma and cold ($\sim10$ K)
molecular gas is nearly a million times greater than that between air and granite.
So while one might naturally expect that the working surface of a jet can drive sound waves and shocks
into the tenuous X-ray atmosphere, it is more difficult to explain the growing
literature reporting observations of massive atomic and molecular outflows
apparently entrained by jets (e.g.,
\citealt{morganti05,morganti13,rupke11,alatalo11,alatalo15, dasyra15, cicone14, cicone18}), or
uplifted in the wakes of the buoyant hot bubbles they inflate (e.g.,
\citealt{mcnamara14,mcnamara16,russell14,russell16a,russell16b,russell17}).
One might instead expect molecular nebulae to act like seawalls, damping turbulence, breaking waves in the hotter phases of the ISM, and redirecting jets.
Recent single-dish and Atacama Large Millimeter/submillimeter Array (ALMA) observations
of cool core clusters nevertheless reveal billions of solar
masses of cold gas in kpc-scale filaments draped around
the rims of radio lobes or X-ray cavities (e.g., Perseus: \citealt{salome08,lim08}, Phoenix:  \citealt{russell16b}, Abell 1795: \citealt{russell17}, M87: \citealt{simionescu18}), or trailing behind them as if drawn upward by their
buoyant ascent (e.g., Abell 1835: \citealt{mcnamara14}; 2A 0335+096: \citealt{vantyghem16}; PKS 0745-191: \citealt{russell16a}).

Such a coupling would be easier to understand were it the manifestation of a
top-down multiphase condensation cascade, wherein both the
warm ionized and cold molecular nebulae are
pools of cooling gas clouds that rain from the ambient hot halo.
The disruption of this halo into a multiphase medium is regulated by the survivability of thermal
instabilities, which lose entropy over a cooling time $t_\mathrm{cool}$,
descend on a free-fall time $t_\mathrm{ff}$, and remain long-lived only if their
local density contrast increases as they sink (e.g., \citealt{voit17a}).
This implies that there is an entropy threshold for the onset of nebular emission
in BCGs, long known to exist observationally \citep{rafferty08,cavagnolo08},
set wherever the cooling time becomes short compared to the effective gas dynamical timescale.
This underlying principle is not
new (e.g., \citealt{hoyle53,rees77,binney77,cowie80,nulsen86,balbus89}), but has
found renewed importance in light of recent  papers arguing that it may be
fundamental to all of galaxy evolution
\citep{pizzolato05,pizzolato10,marinacci10,mccourt12, sharma12,gaspari12,gaspari13,gaspari15,gaspari17b,gaspari18,voit15,voit15d,voit15b,voit15c,voit17a,voit18,li15,prasad15, prasad17, prasad18, singh15, mcnamara16, yang16, meece17, hogan17, main17,pulido18}.

Amid minor disagreement over the importance of the free-fall time,
(compare, e.g., \citealt{voit15d}, \citealt{mcnamara16} and \citealt{gaspari17b}),
these works suggest that the existence of this threshold establishes
a stochastically oscillating but tightly self-regulated
feedback loop between ICM cooling and AGN heating.
The entire process would be mediated by chaotic cold accretion (CCA)
onto the central supermassive black hole \citep{gaspari13}, a prediction
that has recently found observational support with the detection of cold clouds falling toward black hole fuel reservoirs (e.g., \citealt{tremblay16}; Edge et al.~in prep.).
The radio jets that the black hole launches, and the buoyant hot bubbles it inflates,
inject sound waves, shocks, and turbulence into the X-ray bright halo,
lowering the cooling rate and acting as a thermostat for the heating-cooling feedback
loop (e.g., \citealt{birzan04,birzan12,zhuravleva14,hlavacek12,hlavacek15,gaspari17a}). Those same outflows can adiabatically uplift low entropy
gas to an altitude that crosses the thermal instability threshold,
explaining their close spatial assocation with molecular filaments
and star formation \citep{tremblay15,russell17}.
In this scenario, a supermassive black hole acts
much like a mechanical pump in a water fountain\footnote{The supermassive black hole, in this case,
 is akin to the ``pump-like'' action of supernova feedback driving similar fountains in less massive galaxies \citep{fraternali08,marinacci11,marasco13,marasco15}.} (e.g.~\citealt{lim08,salome06,salome11}),
wherein cold gas drains into the black
hole accretion reservoir, powering jets, cavity inflation, and therefore
a plume of low-entropy gas uplifted as they rise.
The velocity of this cold plume is often well below both the escape speed from
the galaxy and the Kepler speed at any given radius (e.g., \citealt{mcnamara16}), and so
those clouds that do not evaporate or form stars should then rain back toward
the galaxy center from which they were lifted.
This, along with merger-induced gas motions \citep{lau17} and the feedback-regulated precipitation of thermal instabilities from the
hot atmosphere, keeps the fountain long-lived and oscillatory.
The apparently violent and bursty cluster core must nevertheless be the engine
of a process that is smooth over long timescales, as the remarkably fine-tuned
thermostatic control of the heating-cooling feedback loop
now appears to persist across at least ten billion years of cosmic time (e.g.,
\citealt{birzan04,birzan08,rafferty06,dunn06,best06,best07,mittal09,dong10,hlavacek12,hlavacek15,webb15,simpson13,mcdonald13b,mcdonald16,mcdonald17,mcdonald18,
 bonaventura17}).

These hypotheses are testable.
Whether it is called ``chaotic cold accretion'' \citep{gaspari13}, ``precipitation'' \citep{voit15d}, or ``stimulated feedback'' \citep{mcnamara16}, the threshold criterion predicts that the
kinematics of the hot, warm, and cold phases of the ISM should retain memory of
their shared journey along what is ultimately the same thermodynamic pathway
\citep{gaspari18}. Observational tests for the onset of nebular emission, star
formation, and AGN activity, and how these may be coupled to this threshold, have
been underway for many years (e.g.,
\citealt{cavagnolo08,rafferty08,sanderson09,tremblay12b,tremblay14,tremblay15,mcnamara16,voit18,hogan17,main17,pulido18}).
The multiphase uplift hypothesis, motivated by theory and simulations \citep{pope10,gaspari12,wagner12,li14a,li14b,li15}, is corroborated by
observations of kpc-scale metal-enriched outflows along the radio axis (e.g., \citealt{simionescu09,kirkpatrick11}), and an increasing number of ionized and
molecular filaments spatially associated with jets or cavities (e.g,
\citealt{salome08,mcnamara14,tremblay15,vantyghem16,russell17}).

More complete tests of these supposed kpc-scale molecular fountains
will require mapping the kinematics
of \textit{all} gas phases in galaxies.
As we await a replacement for the \textit{Hitomi}
mission to reveal the velocity structure of the hot phase \citep{hitomi16,hitomi18, fabian17},
combined ALMA and optical integral field unit (IFU) spectrograph
observations of cool core BCGs can at least begin to
further our joint understanding of the cold molecular and warm ionized gas motions, respectively.
To that end, in this paper we present new
ALMA observations that map the kinematics of cold gas in the Abell 2597 BCG.
We compare these with new Multi-Unit Spectroscopic Explorer (MUSE, \citealt{bacon10}) IFU
data that do the same for the warm ionized phase, as well
as a new deep \textit{Chandra} X-ray image
revealing what is likely filament uplift by A2597's buoyant hot bubbles.
These data are described in \autoref{sec:observations},
presented in \autoref{sec:results}, and discussed in \autoref{sec:discussion}.
Throughout this paper
we assume $H_0 = 70$ km s$^{-1}$ Mpc$^{-1}$, $\Omega_M = 0.27$, and
$\Omega_{\Lambda} = 0.73$.  In this cosmology, 1\arcsec\ corresponds
to 1.549 kpc at the redshift of the A2597 BCG ($z=0.0821$),
where the associated luminosity and
angular size distances are 374.0 and 319.4 Mpc, respectively, and the
age of the Universe is 12.78 Gyr. Unless otherwise noted,
all images are centered on
the nucleus of the A2597 BCG at
Right Ascension (R.A.) 23$^{\mathrm{h}}$ 25$^{\mathrm{m}}$ 19.7$^{\mathrm{s}}$
and Declination  $-12$\arcdeg\ 07\arcmin\ 27\arcsec\ (J2000),
with East left and North up.

\begin{deluxetable*}{cccccc}
 \tabletypesize{\footnotesize}
 \tablecaption{\textsc{Summary of Abell 2597 Observations}}
 \tablehead{
  \colhead{Waveband / Line} &
  \colhead{Facility} &
  \colhead{Instrument / Mode} &
  \colhead{Exp. Time} &
  \colhead{Prog. / Obs. ID (Date)} &
  \colhead{Reference}
 }
 \colnumbers
 \startdata
 \label{tab:observation_summary}
 X-ray (0.2-10 keV) & \textit{Chandra} & ACIS-S & 39.80 ksec & 922 (2000 Jul 28) & \citet{mcnamara01, clarke05} \cr
 \nodata & \nodata & \nodata & 52.20 ksec & 6934 (2006 May 1) & \citet{tremblay12a,tremblay12b} \cr
 \nodata & \nodata & \nodata & 60.10 ksec & 7329 (2006 May 4) & \citet{tremblay12a,tremblay12b} \cr
 \nodata & \nodata & \nodata & 69.39 ksec & 19596 (2017 Oct 8) & Tremblay et al.~(in prep) \cr
 \nodata & \nodata & \nodata & 44.52 ksec & 19597 (2017 Oct 16) & (Large Program 18800649) \cr
 \nodata & \nodata & \nodata & 14.34 ksec & 19598 (2017 Aug 15) & \nodata \cr
 \nodata & \nodata & \nodata & 24.73 ksec & 20626 (2017 Aug 15) & \nodata \cr
 \nodata & \nodata & \nodata & 20.85 ksec & 20627 (2017 Aug 17) & \nodata \cr
 \nodata & \nodata & \nodata & 10.92 ksec & 20628 (2017 Aug 19) & \nodata \cr
 \nodata & \nodata & \nodata & 56.36 ksec & 20629 (2017 Oct 3) & \nodata \cr
 \nodata & \nodata & \nodata & 53.40 ksec & 20805 (2017 Oct 5) & \nodata \cr
 \nodata & \nodata & \nodata & 37.62 ksec & 20806 (2017 Oct 7) & \nodata \cr
 \nodata & \nodata & \nodata & 79.85 ksec & 20811 (2017 Oct 21) & \nodata \cr
 \nodata & \nodata & \nodata & 62.29 ksec & 20817 (2017 Oct 19) & \nodata \cr
 \hline
 Ly$\alpha$ $\lambda$1216 \AA & \textit{HST}     &  STIS F25SRF2    & 1000 sec  & 8107 (2000 Jul 27)  &  \citet{odea04,tremblay15} \cr
 FUV Continuum & \nodata  &  ACS/SBC F150LP  & 8141 sec  & 11131 (2008 Jul 21) &  \citet{oonk10,tremblay15} \cr
 [\ion{O}{2}]$\lambda$3727 \AA  &     \nodata    &  WFPC2 F410M  & 2200 sec & 6717 (1996 Jul 27) & \citet{koekemoer99} \cr
 $B$-band \& [\ion{O}{2}]$\lambda$3727 \AA   & \nodata        & WFPC2 F450W     &  2100 sec    & 6228 (1995 May 07)  & \citet{koekemoer99}  \cr
 $R$-band \& H$\alpha$+[\ion{N}{2}] & \nodata  & WFPC2 F702W & 2100 sec & 6228 (1995 May 07) & \citet{holtzman96} \cr
 H$_2 1-0$ S(3) $\lambda1.9576 \mu$m                     & \nodata              & NICMOS F212N   &   12032 sec     & 7457 (1997 Oct 19)   & \citet{donahue00} \cr
 $H$-band    &  \nodata             & NICMOS F160W     & 384 sec     & 7457  (1997 Dec 03)   & \citet{donahue00} \cr
 H$\alpha$ (Narrowband) &  Baade 6.5m   &   IMACS / MMTF   &   1200 sec   &  (2010 Nov 30)   &   \citet{mcdonald11b, mcdonald11a} \cr
 $i$-band  &  VLT / UT1  & FORS  & 330 sec & 67.A-0597(A)   &  \citet{oonk11} \cr
 Optical Lines \& Continuum & VLT / UT4  & MUSE   & 2700 sec  &  094.A-0859(A)  &   Hamer et al.~(in prep) \cr
 \hline
 NIR (3.6, 4.5, 5.8, 8 $\mu$m)    & \textit{Spitzer}   &   IRAC   & 3600 sec (each)  &  3506  (2005 Nov 24) &   \citet{donahue07}  \cr
 MIR (24, 70, 160 $\mu$m)    & \nodata          & MIPS     & 2160 sec (each)      &   3506  (2005 Jun 18) & \citet{donahue07} \cr
 MIR (70, 100, 160 $\mu$m)   & \textit{Herschel}   & PACS     & 722 sec (each)       &   13421871(18-20) & \citet{edge10phot} \cr
 FIR (250, 350, 500 $\mu$m)  & \nodata          & SPIRE    & 3336 sec (each)      &  (2009 Nov 30)       &  \citet{edge10phot} \cr
 \hline
 CO(2-1)  &  ALMA   &   Band 6 / 213 GHz   &   3 hrs   &   2012.1.00988.S & \citet{tremblay16} \&  this paper \cr
 \hline
 Radio (8.44 GHz)  &   VLA    &   A array    &   15 min    &   AR279  (1992 Nov 30)  &   \citet{sarazin95} \cr
 4.99 GHz          & \nodata   &  A array     & 95 min    &  BT024  (1996 Dec 7)   &  \citet{taylor99, clarke05} \cr
 1.3 GHz           &  \nodata    &  A array    & 323 min    &  BT024 (1996 Dec 7)  &  \citet{taylor99,clarke05} \cr
 330 MHz           &  \nodata    &   A array   &  180 min   &  AC647  (2003 Aug 18) & \citet{clarke05} \cr
 330 MHz           &  \nodata    &  B array    &   138 min   & AC647 (2003 Jun 10)  &  \citet{clarke05} \cr
 \enddata
 \tablecomments{A summary of all Abell 2597 observations used (either directly or indirectly) in this analysis, in decending order
  from short to long wavelength (i.e. from X-ray through radio). (1) Waveband
  or emission line targeted by the listed observation; (2) telescope used; (3) instrument, receiver setup, or array configuration used;
  (4) on-source integration time; (5) facility-specific program or proposal ID (or observation ID in the case of \textit{Chandra}) associated
  with the listed dataset; (6) reference to publication(s) where the listed data first appeared, or were otherwise discussed in detail. Further details for most of these observations,
  including Principal Investigators, can be found in Table 1 of \citet{tremblay12b}.  }
\end{deluxetable*}

\section{Observations \& Data Reduction} \label{sec:observations}

This paper synthesizes a number of new and archival
observations of the A2597 BCG, all of which are summarized
in \autoref{tab:observation_summary}.
Here we primarily describe the new ALMA and MUSE datasets that comprise
the bulk of our analysis. All Python codes / Jupyter Notebooks we have created to enable this analysis
are
publicly available in an online repository\footnote{This code repository is archived at DOI: \href{http://doi.org/10.5281/zenodo.1233825}{10.5281/zenodo.1233825}, and also available at \url{https://github.com/granttremblay/Tremblay2018_Code}.} \citep{papercode}.

\subsection{ALMA CO(2-1) Observations} \label{sec:almareduction}

ALMA observed the Abell 2597 BCG for three hours
across three scheduling
blocks executed between 17-19 November 2013 as part of
Cycle 1 program 2012.1.00988.S (P.I.: Tremblay).
One baseband was centered on the $J=2-1$ rotational line
transition of carbon monoxide ($^{12}$CO) at 213.04685 GHz (rest-frame 230.538001 GHz
at $z=0.0821$).
CO(2-1) serves as a bright tracer for the otherwise
unobservable cold molecular hydrogen gas (H$_2$) fueling star formation
throughout the galaxy (H$_2$ at a few tens of Kelvin is invisible because it lacks a permanent
electric dipole moment).
The other three basebands sampled the local rest-frame $\sim230$ GHz continuum
at 215.0, 227.7, and 229.7 GHz, enabling
continuum subtraction for the CO(2-1) data and an (ultimately unsuccessful)
ancillary search for radio recombination lines.

The ALMA correlator was set to Frequency Division Mode (FDM),
delivering a native spectral (velocity) resolution of 0.488 MHz ($\sim1.3$ km s\mone)
across an 1875 MHz bandwidth per baseband.
Baselines between the array's 29 operational 12 m antenna spanned $17-1284$ m,
delivering a best possible angular resolution at 213 GHz of
$0\farcs37$ within a $\sim 28\arcsec$ primary beam,
easily encompassing the entire galaxy in a single pointing.
In comparing the total recovered ALMA CO(2-1) flux with an older single-dish
IRAM 30m observation \citep{tremblay12b}, we find no evidence that any extended emission has been ``resolved out''
by the interferometer.

Observations of A2597 were bracketed by
slews to Neptune as well as the quasars J2258-2758 and J2331-1556,
enabling amplitude, flux, and phase calibration.
Raw visibilities were imported, flagged, and reduced into calibrated measurement
sets using \texttt{CASA} version 4.2 \citep{mcmullin07}. In addition to applying
the standard phase calibrator solution, we iteratively
performed phase-only self-calibration using the galaxy's own continuum, yielding
a $14\%$ improvement in RMS noise.
We used the \texttt{UVCONTSUB} task to fit and subtract the continuum from the CO(2-1)
spectral window in the $uv$ plane.
We then deconvolved and imaged the continuum-free CO(2-1) measurement set
using the \texttt{CLEAN} algorithm with \texttt{natural} weighting,
improving sensitivity to the filamentary
outskirts of the nebula\footnote{We also experimented with a number of different weighting
 schemes, including \texttt{Briggs} with a \texttt{robust} parameter
 that ranged from -2.0 (roughly \texttt{uniform}) to 2.0 (close to \texttt{natural}).
 We show only \texttt{natural} weighting throughout this paper, partially because
 our results are not strongly dependent on the minor differences between the various
 available algorithms.}.

The final data cube reaches an RMS sensitivity and angular resolution of
$0.16$ mJy beam$^{-1}$ per 40 km~s\mone\
channel with a $0\farcs715\times0\farcs533$ synthesized beam at P.A. = $74^\circ$, enabling us to resolve molecular gas down to physical scales of $\sim 800$ pc.
As indicated in figure captions,
some ALMA images presented in this paper
use Gaussian-weighted $uv$ tapering of the outer baselines
in order to maximize sensitivity to the most extended structures,
expanding the synthesized beam to a size of $0\farcs944 \times 0\farcs764$ at a P.A. of $86^\circ$. The captions
also note whether we have binned the data (in the $uv$ plane) to
5, 10, or 40 km~s\mone
channels, as dictated by sensitivity needs for a given science question.
All CO(2-1) fluxes and linewidths reported in this paper are corrected
for response of the primary
beam (\texttt{pbcor = True}).

We have also created an image of the rest-frame 230 GHz
continuum point source associated with the AGN by summing emission
in the three line-free basebands.
The \texttt{CLEAN} algorithm was set to use \texttt{natural} weighting,
and yielded a continuum map with a synthesized beam
of $0\farcs935 \times 0\farcs747$ at a P.A. of $87^\circ$.
The peak (and therefore total) flux measured from the
continuum point source is  $13.6 \pm 0.2$ mJy at 221.3 GHz,
detected at 425$\sigma$ over the background
RMS noise. It was against this continuum ``backlight''
that \citet{tremblay16} discovered infalling cold molecular clouds seen
in absorption (see \autoref{sec:natureresult}).
We note that the continuum also features $\sim3\sigma$ extended
emission. If one includes this in the flux measurement, it rises
to $14.6 \pm 0.2$ mJy.

This paper also presents CO(2-1) line-of-sight velocity and velocity
dispersion maps made from the ALMA data using the ``masked moment''
technique described by \citet{dame11} and implemented by Timothy Davis\footnote{\url{https://github.com/TimothyADavis/makeplots}}.
The technique takes into account spatial and spectral coherence
in position-velocity space by first smoothing the clean data cube
with a Gaussian kernel whose FWHM is equal to that of the synthesized beam.
The velocity axis is then also smoothed with a guassian, enabling creation
of a three dimensional mask that selects all pixels above a
1.5$\sigma$ flux threshold.
Zeroth, First, and Second moment maps of integrated intensity,
flux-weighted mean velocity, and velocity dispersion (respectively)
were created using this mask on the original (unsmoothed) cube,
recovering as much flux as possible
while suppressing noise.
As we will discuss in \autoref{sec:velocitystructure},
the inner $\sim 10$ kpc of the galaxy contains molecular gas
arranged in two superposed (blue- and redshifted) velocity structures.
We have therefore also created CO(2-1) velocity and velocity dispersion
maps that fit two Gaussians along the same lines of sight. The codes
used to accomplish this are included in the software repository
that accompanies this paper \citep{papercode}.

\subsection{MUSE Optical Integral Field Spectroscopy} \label{sec:musedata}

We also present new spatial and spectral mapping
of optical stellar continuum and nebular emission lines in the A2597 BCG using an observation
from MUSE \citep{bacon10}. MUSE is a high-througput,
wide-FoV, image-slicing integral field unit (IFU) spectrograph mounted at
UT4's Nasmyth B focus on the Very Large Telescope (VLT).
Obtained as part of ESO programme 094.A-0859(A) (PI: Hamer),
this observation was carried out in MUSE's seeing-limited WFM-NOAO-N
configuration on the night of 11 October 2014.
While the $\sim1\arcmin\times1\arcmin$ FoV of MUSE easily covered the entire galaxy
in a single pointing, a three-point dither was used over a $3\times900$ (2700) sec
integration time in order to reduce systematics. Throughout the
observation, the source was at a mean airmass of
$1.026$ with an average $V$-band (DIMM) seeing of $\sim1\farcs2$.

The raw data were reduced using version 1.6.4 of the standard MUSE pipeline \citep{weilbacher14},
automating bias subtraction, wavelength and flux calibration, as well as
illumination-, flat-field, and differential atmospheric diffraction corrections.
In addition to the sky subtraction automated by the pipeline, which
uses a model created from a ``blank sky'' region of the FoV,
we have performed an additional sky subtraction
using a Principal Component Analysis (PCA) code by Bernd Husemann
and the Close AGN Reference Survey\footnote{\url{http://www.cars-survey.org}}
(CARS; \citealt{husemann16, husemann17}).
We have also corrected the datacube for Galactic foreground
extinction using $A_V=0.082$, estimated from the \citet{schlafly11}
recalibration of the \citet{schlegel98} \textit{IRAS}+\textit{COBE}
Milky Way dust map assuming $R_V=3.1$.

The final MUSE datacube maps the entire galaxy
between $4750~$\AA $~< \lambda < $  $~9300$ \AA\ with a spectral
resolution of  $\sim2.5$ \AA. The FWHM of its seeing-limited point-spread
function, sampled with $0\farcs2$ pixels, is $1\farcs0$ and $0\farcs8$ on the bluest and reddest
ends of the spectral axis, respectively. This is close to
the spatial resolution of our ALMA CO(2-1) map, enabling comparison of
the kinematics and morphology of warm ionized and cold molecular gas phases
on nearly matching spatial scales.

In pursuit of that goal, we have created a number of higher level MUSE data
products by decoupling and modeling the stellar and nebular components of the galaxy
with \textsc{PyParadise},
also used by the CARS team as part of their custom MUSE
analysis tools \citep{walcher15,husemann16,weaver18}.
\textsc{PyParadise} iteratively performs non-negative linear least-squares fitting of
stellar population synthesis templates to the stellar spectrum of every relevant
spectral pixel (``spaxel'')
in the MUSE cube, while independently finding the best-fit line-of-sight
velocity distribution with a  Markov Chain Monte Carlo (MCMC) method.
The best-fit stellar spectrum is then subtracted from each spaxel,
yielding residuals that contain nebular emission lines. These are fit with a
linked chain of Gaussians that share a common radial velocity, velocity dispersion,
and priors on expected emission line ratios (e.g., the line ratios of the [\ion{O}{3}] and [\ion{N}{2}] doublets are fixed to 1:3).
Uncertainties on all best-fit stellar and nebular parameters
are then estimated using a Monte-Carlo bootstrap approach wherein
both continuum and emission lines
are re-fit 100 times as the spectrum is
randomly modulated within the error of each spaxel.

While the nebular emission lines in the A2597 MUSE observation were bright
enough to be fit at the native (seeing-limited) spatial resolution,
the S/N of the stellar continuum was low enough to necessitate spatial binning.
We have applied the Voronoi tesselation technique using a Python code kindly provided\footnote{\url{http://www-astro.physics.ox.ac.uk/~mxc/software/\#binning}}
by Michele Cappellari \citep{cappellari03}.
The MUSE cube was tessellated to achieve a minimum S/N of 20 (per bin)
in the line-free stellar continuum.

The products from \textsc{PyParadise} then enabled creation
of spatially resolved flux, velocity, and velocity dispersion maps
of those emission lines most relevant to our study, namely H$\alpha$,
[\ion{O}{1}] $\lambda$6300 \AA,  [\ion{O}{3}] $\lambda$5007 \AA, and
H$\beta$, along with Voronoi-binned velocity and FWHM maps for the galaxy's stellar
component. We have also created Balmer decrement (H$\alpha$ / H$\beta$ ratio),
color excess ($E(B-V)$), and optical extinction ($A_V$)
maps by dividing the H$\alpha$ and H$\beta$ maps
and scaling the result by following equation (1)
in \citet{tremblay10}. Finally, we show an
electron density map made by scaling
the ratio of forbidden sulfur lines
(i.e., [\ion{S}{2}]$\lambda\lambda$ 6717 \AA\ / 6732 \AA; \citealt{osterbrock06}) using
using the calibration of \citealt{proxauf14} (see their Eq.~3)
and assuming an electron temperature of $T_e=10^{4}$ K.
We repeated this process to make Balmer decrement and electron density maps
from a cube whose spaxels were binned $4\times4$, increasing signal in the fainter lines. Comparing these maps
to their unbinned counterparts revealed no quantitative difference. We therefore only show the unbinned,
higher spatial resolution maps in this paper.

\subsection{ALMA and MUSE Line Ratio Maps} \label{sec:ratiomaps}

We have also created H$\alpha$/CO(2-1) flux, velocity,
and velocity dispersion ratio maps by dividing the ALMA ``masked moment''
maps from the corresponding MUSE maps. To accomplish this,
we made small WCS shifts in the MUSE maps to match the ALMA CO(2-1) image with
the \textsc{PyRAF} \texttt{imshift} and \texttt{wcscopy} tasks, assuming that
the CO(2-1) and H$\alpha$ photocentroids in the galaxy center as well as a bright, clearly detected ($\gae10\sigma$) ``blob'' of emission to the northwest in both datasets should be aligned.
The needed shifts were minor, and applying them also aligned enough morphologically matching features that we are
confident that the alignment is ``correct'', at least to
an uncertainty that is smaller than the PSF of either observation.
We then confirmed that the ALMA synthesized beam closely matched
the MUSE PSF at  H$\alpha$ (7101 \AA\ and 6563 \AA\ in the observed and rest-frames,
respectively), making smoothing unnecessary.
We then resampled the ALMA data onto the MUSE maps' pixel grids in Python
using \texttt{reproject},
an \texttt{Astropy} affiliated package\footnote{\url{https://reproject.readthedocs.io/en/stable/}}.
Depending on science application, the
reprojected ALMA image was then either divided directly from the MUSE map,
or divided after normalization or rescaling by some other factor
(for example, to convert pixel units).
The Python code used to create these maps, along with all MUSE
and ALMA data products, is included
in this paper's software repository \citep{papercode}.

\begin{figure}
 \begin{center}
  \includegraphics[scale=0.48]{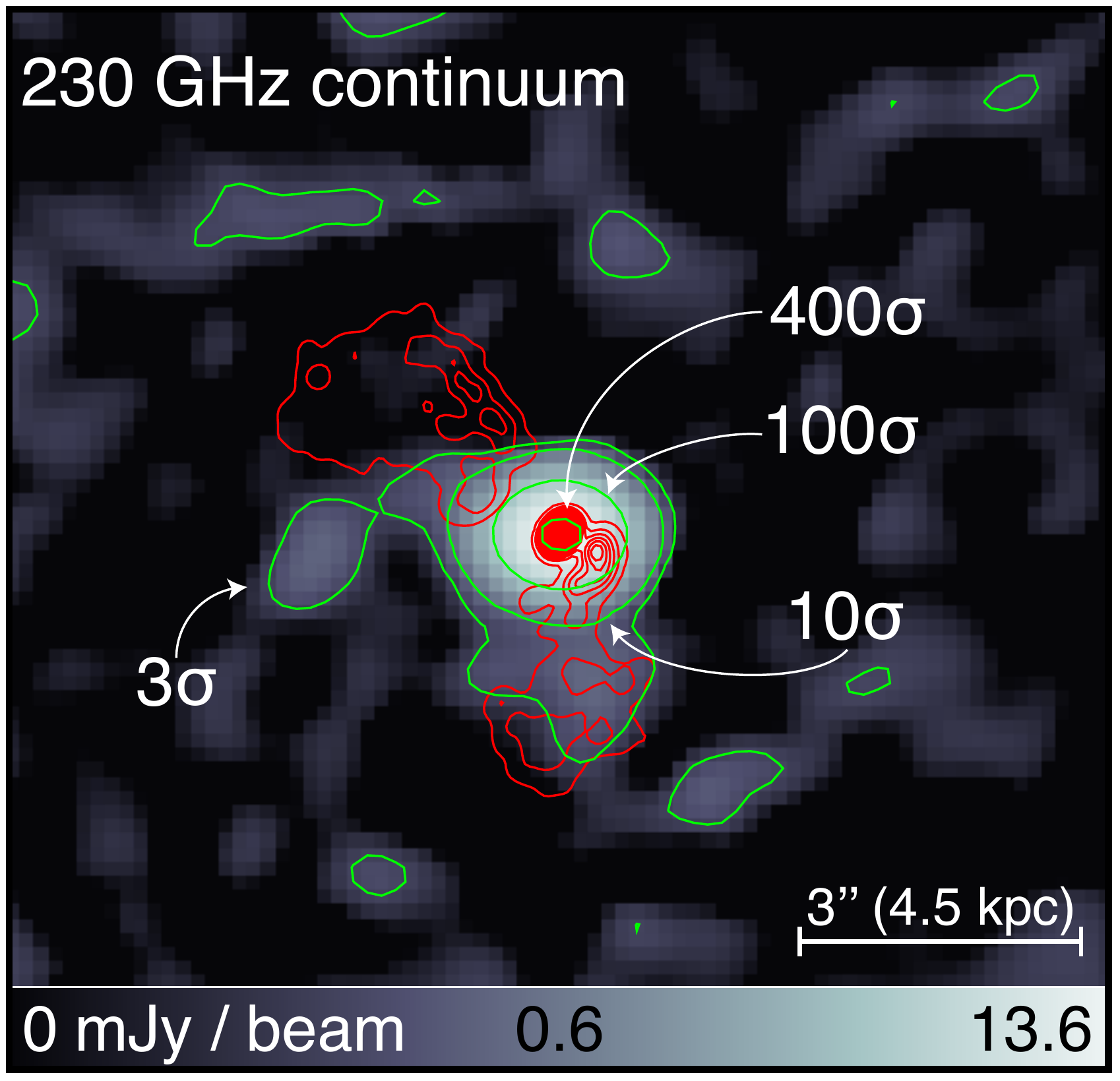}
 \end{center}
 \vspace*{-5mm}
 \caption{The ALMA 230 GHz continuum signal, summed over three basebands
  redward of the CO(2-1) line. The map is dominated by a
  mm synchrotron continuum point source associated with the AGN at the galaxy center, with a
  flux density of $13.6 \pm 0.2$ mJy.
  Contours marking the 8.4 GHz VLA observation of the compact steep spectrum radio source are overlaid in red. The $10\sigma$ contour is consistent with an unresolved point source.
  A log stretch has been applied to the
 data so as to best show the $3\sigma$ extended emission against the $\gae 400\sigma$ point source.
 Much of this extended emission is likely to be noise, though the extension to the
 south along the 8.4 GHz radio source may be real.
 We are unlikely to have detected any extended dust continuum emission, given
 the FIR fluxes shown in \autoref{fig:SED}.  }
 \label{fig:continuum}
\end{figure}

\subsection{Adoption of a systemic velocity}

All ALMA and MUSE velocity maps shown in this paper are projected
about a zero-point that is set to the stellar systemic
velocity of the A2597 BCG at $z=0.0821 \pm 0.0001$ ($cz = 24,613 \pm 29$ km s\mone).
As discussed in the Methods section of \citet{tremblay16},
this velocity is consistent with \ion{Ca}{2}~\textsc{h+k} and G-band
absorption features tracing the galaxy's stellar component,
a cross-correlation of galaxy template spectra with
all major optical emission and absorption lines in the galaxy \citep{voit97,koekemoer99,taylor99},
an \ion{H}{1} absorption feature \citep{odea94}, and the ALMA CO(2-1) emission
line peak itself \citep{tremblay16}.
It is, therefore, the best-known systemic velocity for the system,
within $\sim 60$ km s\mone.

\subsection{Deep Chandra X-ray data} \label{sec:archivaldata}

Finally, we have combined all available \textit{Chandra X-ray Observatory}
data for A2597, spanning 626.37 ksec in total integration time
across fourteen separate ACIS-S observations.
The oldest three of these (see \autoref{tab:observation_summary})
were previously published
(ObsID 922, PI: McNamara and ObsIDs 6934 and 7329, PI: Clarke; \citealt{mcnamara01, clarke05,tremblay12a,tremblay12b}), while the
latest eleven were recently observed as part of Cycle 18 Large Program
18800649 (PI: Tremblay). This new dataset will be analyzed in detail
by Tremblay et al.~(in prep). Here, we show only the deep image
for the purposes of comparing it with the ALMA and MUSE data.

To create this deep image, all fourteen ACIS-S observations were
(re)-reduced, merged, and exposure corrected using \textsc{ciao} version 4.9 \citep{fruscione06}
with version 4.7.5.1 of the Calibration Database. All exposures centered
the cluster core (and therefore the BCG) on the nominal aimpoint
of the back-illuminated S3 chip.
We have applied a radially varying gradient filter to the final merged \textit{Chandra}
image using a Gaussian Gradient Magnitude (GGM)
technique recently implemented to highlight surface brightness edges in \textit{Chandra} data by \citet{sanders16b,sanders16a,walker17}.
The codes we used to accomplish this have been kindly provided by Jeremy Sanders, and are publicly available\footnote{\url{https://github.com/jeremysanders/ggm}}.

\begin{figure}
 \begin{center}
  \includegraphics[width=0.485\textwidth]{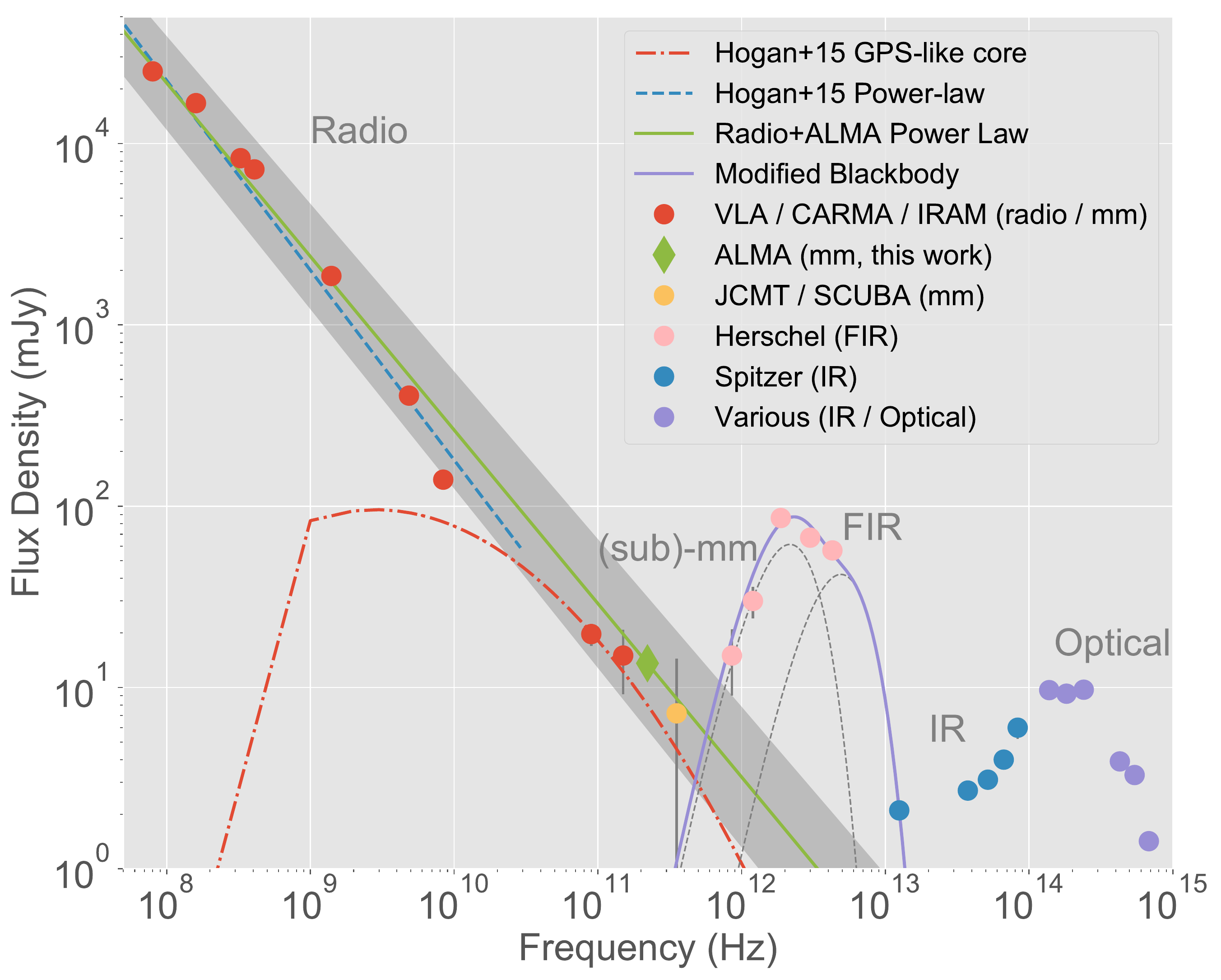}
 \end{center}
 \vspace*{-5mm}
 \caption{Radio-through-optical SED for Abell 2597, including the new ALMA  mm continuum point. Dashed and solid lines show various fits to components of the spectrum
 including a one- and two-component fit to the radio and ALMA data \citep{hogan15b,hogan15a}, as well
 as a modified blackbody fit to the far-infrared \textit{Herschel} data \citep{mittal11,mittal12}.
 Observation details (including dates) and references for all photometric points are given in
 \autoref{tab:observation_summary}. Errorbars are shown on the plot, though in many cases remain invisible because they are smaller than the data point. The gray shaded region shows
 the error on the single powerlaw fit to both the radio and ALMA continuum data.
 These fits are discussed in \autoref{sec:natureresult}.
 }
 \label{fig:SED}
\end{figure}

\begin{figure*}
 \begin{center}
  \includegraphics[width=\textwidth]{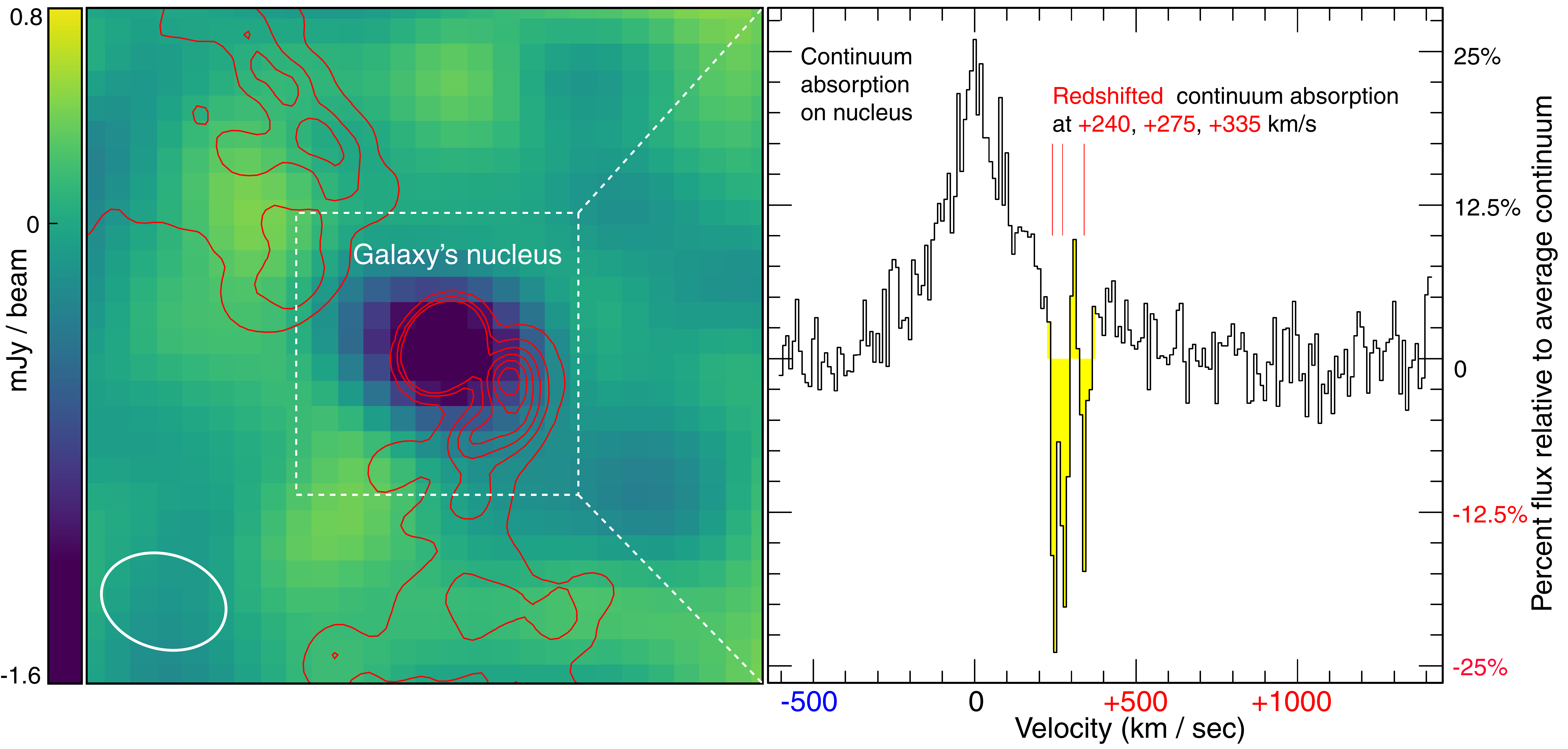}
 \end{center}
 \vspace*{-3mm}
 \caption{A summary of the primary result from \citet{tremblay16},
  showing three compact ($\lae 40$ pc) molecular clouds moving deeper
  into the galaxy and toward its nucleus at $\sim +300$ \kms. The clouds are likely
  in close proximity (within $\sim 100$ pc) to the central supermassive black hole,
  and therefore may play a direct role in fueling the black hole's accretion reservoir.
  (\textit{left}) A slice through the continuum-subtracted ALMA
  CO(2-1) datacube, 10 \kms\ in width and centered on $+240$ \kms\ relative to the
  galaxy's systemic velocity. A region of ``negative emission'', arising from
  continuum absorption, appears as a
  dark spot the size
  of the ALMA beam, whose $0\farcs715 \times 0\farcs533$ ($\sim 1$ kpc $\times \sim0.8$ kpc)
  size is indicated by the white ellipse in the bottom left corner.
  8.4 GHz radio contours are shown in red.
  The innermost contours of the radio core associated with the AGN have been
  removed to aid viewing
  of the ALMA continuum absorption feature.
  Extracting the CO(2-1) spectrum from a region bounding the galaxy's nucleus
  (roughly marked by the dashed white box) reveals the spectrum in the rightmost panel (adapted from \citealt{tremblay16}).
 }
 \label{fig:nature}
\end{figure*}

\section{Results} \label{sec:results}

\subsection{``Shadows'' cast by inflowing cold clouds} \label{sec:natureresult}

The ALMA observation is dominated by a bright continuum point source,
shown in \autoref{fig:continuum}.
Its flux at 221.3 GHz is $13.6 \pm 0.2$ mJy, which we
show as part of a radio-through-optical SED
in \autoref{fig:SED}.
The green line
shows a single powerlaw fit to the
radio and ALMA data points with a spectral index
of $\alpha=0.95 \pm 0.03$ if $S \propto \nu^{-\alpha}$, where $S$ is flux density
and $\nu$ is frequency.
The surrounding gray region shows the error on that fit,
and entirely encompasses the two-component radio-only fit
by \citet{hogan15a}. That model is shown in blue dashed
and red dash-dotted lines and includes, respectively, a
powerlaw with spectral index $\alpha = 1.18 \pm 0.06$ and a likely highly variable,
flatter, GPS-like core (see discussion in \citealt{hogan15b,hogan15a}).
Some curvature in the radio spectrum is evident, though it may be partly artificial
as these data points were collected over the course of more than twenty years, during which time
the source likely varied in brightness.
Regardless, within errors, the new ALMA
data point is consistent with both the single powerlaw and
two-component models, and so it is likely that
the 230 GHz continuum source detected by ALMA is simply
the millimeter tail of the synchrotron continuum entirely
associated with the AGN.

This continuum source acts as a bright backlight cast by the radio jet's
launch site, in close proximity to the $\sim3\times10^8$ \Msol\
black hole in the galaxy center \citep{tremblay12b}.
Against this backlight we found three deep, narrow continuum absorption
features (\autoref{fig:nature}), which we discuss in \citet{tremblay16}.
We suggest that these are ``shadows'' cast
by inflowing cold molecular clouds eclipsing our line of sight to the black hole.
Assuming they are in virial equilibrium, we calculate that the
clouds, whose linewidths are not more than $\sigma_v \lae 6$ km s\mone,
must have sizes no greater than $\sim 40$ pc and masses on the order of
$\sim10^5-10^6$ \Msol, similar to giant molecular clouds in the Milky Way (e.g., \citealt{larson81,solomon87}).
If they are in pressure equilibrium with their ambient multiphase environment,
their column densities must be on the order of $N_\mathrm{H_2} \approx 10^{22-24}$
cm\mtwo.
A simple argument based on geometry and probability, along with corroborating
evidence from the Very Long Baseline Array (VLBA), suggests that these
inflowing cold molecular clouds are within
$\sim100$ pc of the black hole, and falling ever closer toward it \citep{tremblay16}.
These clouds may therefore provide a substantial cold molecular mass flux
to the black hole accretion reservoir, contrary to what might be expected
in a ``hot mode'' Bondi-like accretion scenario.
Regardless, these results establish that some cold molecular gas is clearly
moving inward toward the galaxy center. The remainder of this paper
connects this inflowing gas to the larger galaxy of which it is a part.

\subsection{Morphology of the cold molecular nebula}

\begin{figure*}
 \begin{center}
  \includegraphics[width=\textwidth]{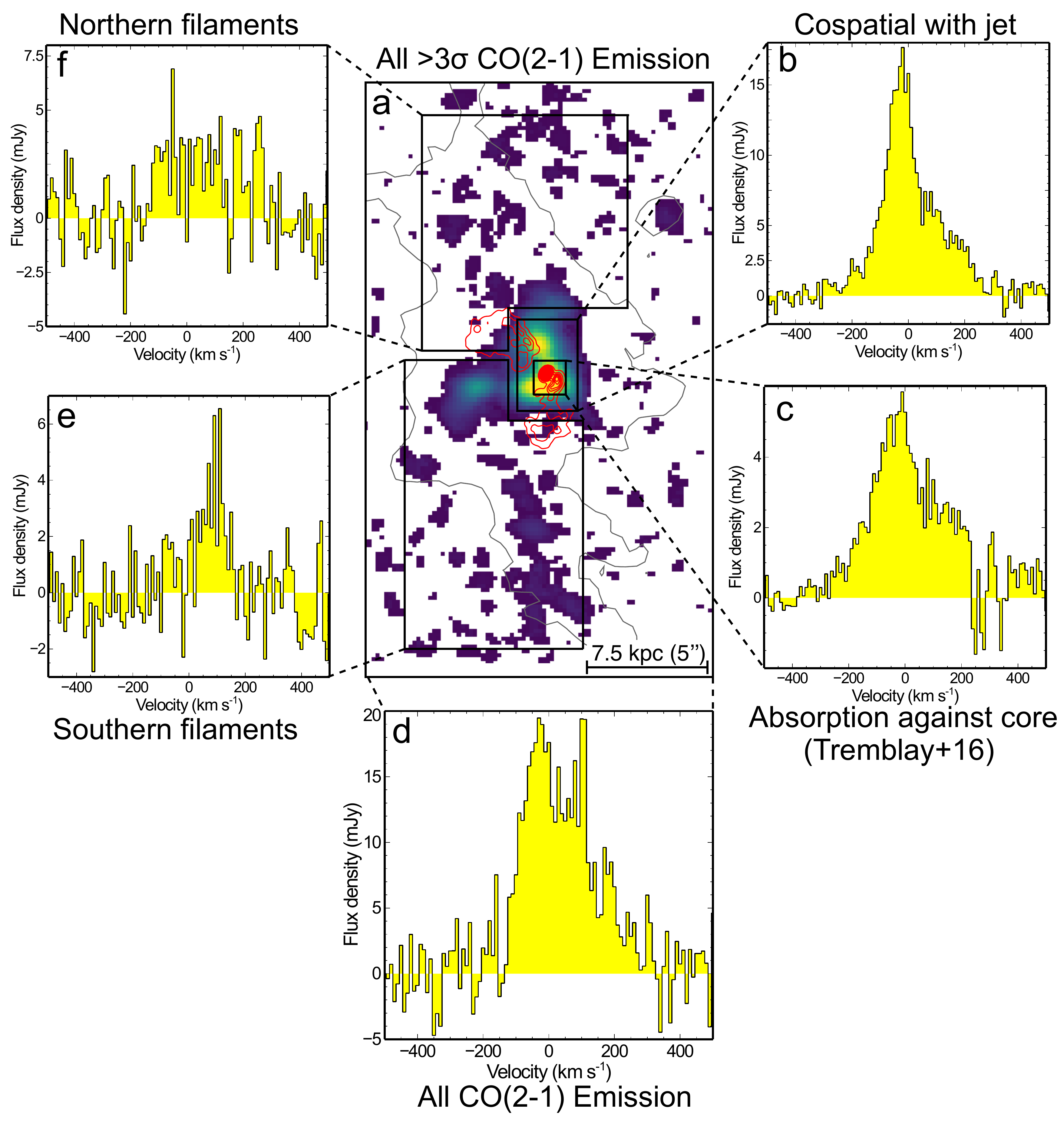}
  \vspace*{-8mm}
 \end{center}
 \caption{An overview of the morphological and spectral characteristics of the ALMA CO(2-1) observation we discuss
  at length in this paper. The central panel (\textit{a}) shows a clipped moment zero (flux) image of all $\ge3\sigma$ CO(2-1)
  emission in the A2597 BCG.
  The various clumps seen likely represent $\gae3\sigma$ peaks of a smoother, fainter distribution of gas below the sensitivity threshold (although some clumps may indeed be discrete).
  For reference, the outer contour of the H$\alpha$ nebula is shown with a solid gray contour. Various apertures are shown in black
  polygons, indicating the (rough) spectral extraction regions
  for the CO(2-1) line profiles shown in the surrounding panels.
  All data are binned to  10 km s\mone\ channels.
  (\textit{b}) The CO(2-1) line profile from a region
  cospatial with the $\sim10$ kpc-scale CSS radio source (red contours on panel \textit{a}). (\textit{c}) An extraction from the nucleus of the galaxy,
  cospatial with the mm and radio core, as well as the stellar isophotal centroid.
  The deep absorption features are discussed in \autoref{sec:natureresult} and \citet{tremblay16}.
  (\textit{d}) All detected emission across the entire nebula. It is this spectrum
  from which we estimate the total gas mass in \autoref{sec:gasmass}.
  Panels (\textit{e}) and (\textit{f}) show the spectra
  extracted from what we call the southern and northern filaments, respectively.
 }
 \label{fig:region_plots}
\end{figure*}

The continuum-subtracted ALMA CO(2-1) data
reveal a filamentary molecular nebula
whose largest angular extent spans the inner 30 kpc (20\arcsec)
of the galaxy (\autoref{fig:region_plots}\textit{a}).
The brightest CO(2-1) emission is
cospatial with the galaxy nucleus, forming a ``V'' shape with an axis
of symmetry that is roughly aligned with the galaxy's stellar minor axis.
In projection, a 12 kpc (8\arcsec) linear filament appears to connect
with the southeastern
edge of the ``V'' and arcs southward. Fainter clumps and filaments,
many of which are part of a smoother distribution of gas just below
the $\ge3\sigma$ clipping threshold shown in \autoref{fig:region_plots}\textit{a},
are found just to the north of the ``V''.

This cold molecular nebula is forming stars across its entire
detected extent, at an integrated rate of $\sim5$ \Msol\ yr\mone\
as measured with a number of observations, including \textit{Herschel} photometry \citep{edge10phot,edge10spec,tremblay12b}.
We have smoothed the \textit{HST}/ACS SBC FUV continuum map
from \citet{oonk11}
with a Gaussian whose FWHM matches that of the synthesized beam in our ALMA
map of integrated CO(2-1) intensity, normalized their surface brightness
peaks, and then divided one map by the other. The quotient
map is close to unity across the nebula, indicating that the star formation rate
surface density (even as traced by extinction-sensitive FUV continuum) is proportional
to the underlying CO(2-1) surface brightness\footnote{This is unsurprising in the context
of a simple \citet{kennicutt98} scenario. It is, however, also important to consider
this result alongside the several known CC BCG filament systems that are clearly \textit{not}
forming stars. A famous example is found in the Perseus/NGC 1275 optical nebula. Many
of its filaments are rich in molecular gas \citep{salome11}, yet largely devoid
of any ongoing star formation (e.g., \citealt{conselice01,canning14}).}.

Where they overlap, the MUSE/ALMA H$\alpha$-to-CO(2-1)
surface brightness ratio map is similarly smooth (see \autoref{sec:musealma}).
Matching H$\alpha$ and CO(2-1) morphology is consistent with the
hypothesis that the optical and mm emission
arises from the same population of clouds, as we will discuss
in \autoref{sec:muse} and \autoref{sec:discussion}.
In \autoref{fig:region_plots}\textit{a}, we show the CO(2-1) emission
bounded by a gray contour that marks the outer extent of the H$\alpha$
emission.
That the molecular nebula appears smaller in angular extent than
the warm ionized nebula is more likely due to a sensitivity floor
than a true absence of cold gas at larger radii.
The ALMA observations do reveal faint, smooth emission in the northern
and southern locales of the warm ionized filaments, though much of it is
simply below the threshold we apply to all CO(2-1) maps
presented in this paper.
That we have detected at least \textit{some} faint molecular emission
in the outer extents of the warm nebula suggests that, were we to observe
to greater depths with ALMA, we might detect CO(2-1) across its entire
extent. This isn't guaranteed, as warm ionized
gas can be present without cold molecular gas
(e.g., \citealt{simionescu18}).
We do note that most ALMA observations of CC BCGs published thus far
generally show molecular filaments cospatial with warm ionized
counterparts \citep{mcnamara14,russell14,vantyghem16,russell16a,russell16b,russell17}.
This has been known long prior to the first ALMA observations, too (see, e.g.,
the single-dish observations of the Perseus filaments
by \citealt{lim08,salome11}).

\begin{figure*}
 \begin{center}
  \includegraphics[scale=0.39]{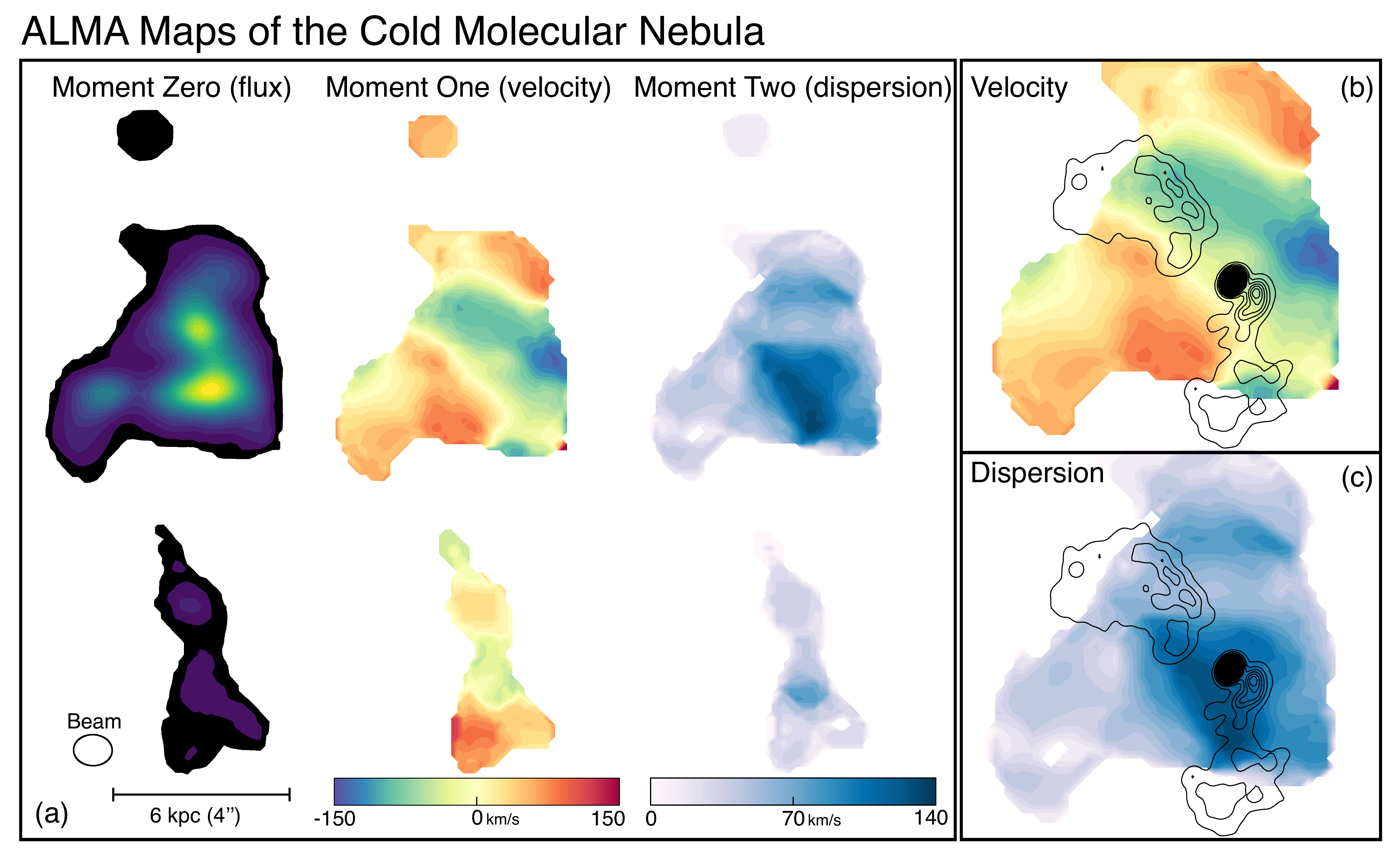}
 \end{center}
 \vspace*{-5mm}
 \caption{(\textit{a}) Zeroth, First, and Second moment maps of
  integrated CO(2-1) intensity, mean velocity, and
  velocity dispersion (respectively) in the cold
  molecular nebula. The maps have been created
  from the ALMA cube
  using the ``masked moment'' technique to preserve
  spatial and spectral coherence of $\ge3\sigma$ structures
  in position-velocity space, as
  described in \autoref{sec:almareduction}. Panels (\textit{b}) and (\textit{c}) show a zoom-in on the nuclear region in the velocity and velocity dispersion maps, respectively.  Take caution when interpreting
  these,
  because there are two velocity components (one approaching/blueshifted, the other receding/redshifted)
  superposed on one another. The velocity structure here is therefore best
  represented by a double-Gaussian fit, which we show in \autoref{fig:fountain_overview}.   }
 \label{fig:alma_momentmaps}
\end{figure*}

\begin{figure*}
 \begin{center}
  \includegraphics[width=\textwidth]{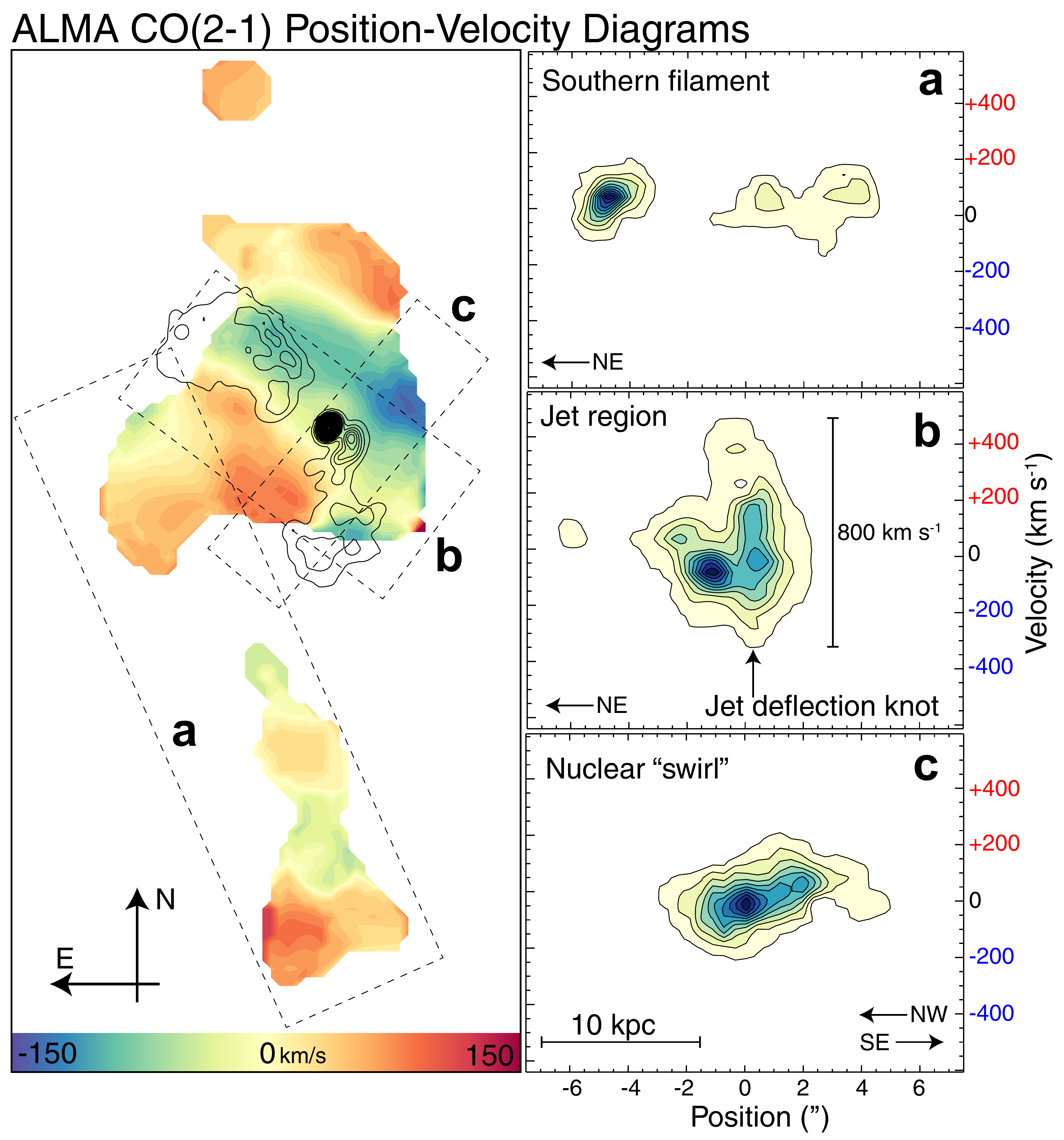}
 \end{center}
 \vspace*{-3mm}
 \caption{Position-velocity (PV) diagrams extracted from the three regions of the molecular nebula.
  The lefthand panel shows the Moment One velocity map from \autoref{fig:alma_momentmaps}, with three position-velocity extraction apertures
  overlaid. The righthand panels show the PV diagrams extracted from these apertures. Arrows are used to show the cardinal orientation of each aperture's long axis  (the slit orientation for panel \textit{c} is roughly perpendicular to that for panels \textit{a} and \textit{b}, and so the relative orientations are admittedly confusing at first glance). Note that while the length of the extraction aperture varied, all diagrams are shown on the spatial same scale in the righthand panels, enabling cross-comparison. Panel \textit{a} shows that the southern filament has a narrow velocity width across its entire length, and no coherent velocity gradient. Panel \textit{b} reveals the broadest velocity distribution of molecular gas in the entire nebula, and includes the region in which the 8.4 GHz radio source bends in position angle, likely because of deflection. Panel \textit{c} shows rotation of molecular gas about the nucleus. All emission shown is $\ge3\sigma$. }
 \label{fig:pv_figure}
\end{figure*}

\subsection{Total mass and mass distribution of the molecular gas} \label{sec:gasmass}

Assuming a CO(2-1) to CO(1-0) flux density ratio of 3.2 \citep{braine92},
we can estimate the total mass of molecular H$_2$ in the nebula following
the relation reviewed by \citealt{bolatto13b}:
\begin{align}
 M_\mathrm{mol} & = \left(\frac{1.05 \times 10^4}{3.2}\right) ~ \left( \frac{X_{\mathrm{CO}}}{X_{\mathrm{CO,~MW}}} \right)                                                                    \\
                & \times \left( \frac{1}{1+z}\right) \left(\frac{S_{\mathrm{CO}}\Delta v}{\mathrm{Jy~km~s}^{-1}}\right)   \left(\frac{D_\mathrm{L}}{\mathrm{Mpc}}\right)^2 M_\odot, \nonumber
 \label{eqn:mass}
\end{align}
where $S_{\mathrm{CO}}\Delta v$ is the integrated CO(2-1) intensity,
$z$ is the galaxy redshift ($z=0.0821$), and $D_L$ its luminosity distance (374 Mpc in our adopted cosmology).
The dominant source of uncertainty in this estimate is
the CO-to-H$_2$ conversion factor $X_{\mathrm{CO}}$
(see, e.g., \citealt{bolatto13b}).
Here we adopt the average value for the disk of the Milky Way
of $X_{\mathrm{CO}} = X_{\mathrm{CO,~MW}} = 2 \times 10^{20}$ cm\mtwo\ $\left(\mathrm{K~km~s}^{-1}\right)^{-1}$.
There is a $\sim30\%$ scatter about this value \citep{solomon87},
minor in comparison to the overriding uncertainty as to the appropriateness
of assuming that the A2597 BCG is at all like the Milky Way.
The true value of the conversion factor depends on
gas metallicity and whether or not the CO emission is optically thick.
The metal abundance of the hot X-ray plasma is $\sim0.5-0.8$ Solar in the inner
$\sim50$ kpc of the A2597 BCG \citep{tremblay12a}, and the velocity
dispersion of individual molecular clouds in the galaxy are similar
to those in the Milky Way \citep{tremblay16}.
Echoing arguments made for the A1835, A1664, and A1795
BCGs in \citet{mcnamara14}, \citet{russell14} and \citet{russell17}, respectively,
we have no evidence to suggest that
the ``true'' $X_{\mathrm{CO}}$ in A2597 should be wildly diffferent from the Milky Way,
as it can often be in ULIRGS \citep{bolatto13b}.
Indeed, \citet{vantyghem17} report one of the first
detections of $^{13}$CO(3-2) in a BCG (RX J0821+0752), and
in doing so find a CO-to-H$_2$ conversion factor
that is only a factor of two lower than that for the Milky Way.
Adopting $X_{\mathrm{CO,~MW}}$ is therefore likely to be the most reasonable choice,
with the caveat that we may be overestimating the total mass by a factor of a few.
This should be taken as the overriding uncertainty
on all mass estimates quoted in this paper.

We fit a single Gaussian to the CO(2-1) spectrum
extracted from a polygonal aperture encompassing all $\ge3\sigma$ emission
in the primary beam corrected cube, binned to 10 km s\mone\ channels (this spectrum is shown in \autoref{fig:region_plots}\textit{d}).
This gives an emission integral of
$S_{\mathrm{CO}}\Delta v = 7.8\pm0.3$ Jy km s\mone\ with a line FWHM of $252 \pm 16$ km s\mone,
which, noting the caveats discussed above, converts to an
H$_2$ gas mass of $M_{\mathrm{H}_2}  = \left(3.2 \pm 0.1\right) \times 10^9$ \Msol.
Within errors, we obtain the same integral for cubes binned to 20 or 40 km s\mone,
and an identical flux with an analytic integral of the line (e.g. adding all $\ge3\sigma$ flux in the cube, rather than fitting a Gaussian).
This mass estimate is a factor of $\sim1.8$ higher
than that in \citealt{tremblay16} because their Gaussian was fit
from $-500$ to $+500$ km s\mone, while ours is fit between $-600$ and $+600$ km s\mone.
This apparently minor difference gives rise to a significant offset
because the former fit misses real emission blueward and redward of the line,
biasing the continuum zero point upward. \citet{tremblay16} therefore
slightly underestimate the total flux, though not to a degree that affects
any of the results reported in that work.

Indeed, factor of two variations in the total mass estimate
do not significantly impact the conclusions drawn in either paper,
especially considering the larger uncertainty coupled to
our assumption for $X_{\mathrm{CO}}$ and the CO(2-1) to CO(1-0)
flux density ratio. It is sufficient
for our purposes to say that the total cold molecular
gas mass in the A2597 BCG is a few billion solar masses.
Given the
critical density of CO(2-1), any reasonable assumption for the three-dimensional volume of the nebula, and the total amount of
cold gas available to fill it, the volume filling
factor of the cold molecular clouds cannot be more than a few percent (\citealt{tremblay16}; see also \citealt{david14, anderson17,temi18}).
Far from a monolithic slab, the cold gas is instead
more like a ``mist'' of many smaller individual clouds and
filaments seen in projection (e.g., \citealt{jaffe01,jaffe05,wilman06,emonts13,mccourt18}).

A significant fraction of the total mass in this ``mist'' is found far from the galaxy's nucleus.
In \autoref{fig:region_plots} we divide the nebula into three primary
components consisting of the bright nuclear region cospatial with the 8.4 GHz radio source (panel \textit{b}),
the northern filaments (panel \textit{f}), and the southern filaments (panel \textit{e}).
Fitting the CO(2-1) spectra extracted from each of these components shows that their
rough fractional contribution to the total
gas mass (i.e., panel \textit{d}) is $\sim70\%$, $\sim10\%$, and $\sim20\%$,
respectively.
This means that although most ($\sim 2.2\times10^9$ \Msol) of the cold gas
is found in the innermost $\sim8$ kpc of the galaxy, $\sim1$ billion \Msol\
of it lies at distances greater than 10 kpc from the galactic center.

\subsection{Velocity structure of the molecular gas} \label{sec:velocitystructure}

In \autoref{fig:alma_momentmaps} we show the
``masked moment'' maps of integrated CO(2-1) intensity, flux-weighted velocity,
and velocity dispersion.
The cold molecular nebula features
complex velocity structure across its spatial extent,
with gas found at projected line-of-sight velocities that
span $\gae300$ km s\mone, arranged roughly symmetrically about the systemic velocity of the galaxy.
Aside from a possible $\pm 100$ km s\mone\ rotation (or ``swirl'') of gas near  the
nucleus (\autoref{fig:alma_momentmaps}\textit{b}, see the blue- and redshifted components to the NW and SE of the radio core, respectively), most of the nebula appears removed from a state of dynamical equilibrium, and poorly mixed (in phase space) with the galaxy's stars. \textit{Almost} everywhere, projected line of sight velocities are below the circular
speed at any given radius, and well below the galaxy's escape velocity.
The kinematics of the molecular nebula can therefore be considered rather
slow, unless most gas motions are contained in the plane of the sky.
This is unlikely, given several recent papers reporting similarly
slow cold gas motions in CC BCGs \citep{mcnamara14,russell14,russell16a,russell16b,russell17,vantyghem16}.
The overall picture for A2597, then, is that of a slow, churning ``mist'' of cold gas, drifting
in the turbulent velocity field of the hot atmosphere,
with complex inward and outward streaming motions.
In the below sections we will argue that these motions are largely
induced by mechanical feedback from the central supermassive black hole,
mediated either by the jets that it launches, or the buoyant X-ray cavities
that those jets inflate.

\subsubsection{Uplift of the Southern Filament}
\label{sec:filament}
The velocity and velocity dispersion maps in \autoref{fig:alma_momentmaps}\textit{a} (center and right)
show largely quiescent structure along the southern filament,
with no monotonic
or coherent gradient in either across its $\sim12$ kpc projected length.
In \autoref{fig:pv_figure}\textit{a} we show a position-velocity (hereafter ``PV'') diagram
of emission extracted from a rectangular aperture around the filament.
The structure is brightest at its northern terminus (i.e., the left-hand side of \autoref{fig:pv_figure}\textit{a}), which serves as the
easternmost vertex of the bright central ``V'' feature
around the galaxy nucleus. Southward from this bright knot, toward the right-hand side of \autoref{fig:pv_figure}\textit{a}, the filament is roughly constant in velocity
centroid and width ($+50-100$ km s\mone and $\sim80-100$ km s\mone, respectively). $\sim6\arcsec$ ($\sim9$ kpc) south of the northern terminus,
however, the filament broadens in velocity dispersion.
Here, near the filament's apex in galactocentric
altitude, it features its largest observed line-of-sight velocity width ($\sim 300$ km s\mone),
with a centroid that is roughly the same as that along its entire length.

The southern filament's velocity structure is
inconsistent with gravitational free-fall \citep{lim08}.
Its projected
length spans $\sim12$ kpc in galactocentric altitude, along which one would
expect a radial gradient in Kepler speed. Its major axis is roughly
parallel (within $\sim20^\circ$) to the projected stellar isophotal minor
axis, but the filament itself  is offset at least 5 kpc to the southeast.
In response to the gravitational potential, gas at high altitude
will have a higher velocity toward the galaxy's nucleus
than it will at its orbital apoapse \citep{lim08}.
It therefore spends a longer amount of time around its high altitude
``turning point'' than it does in proximity to the nucleus.
This is consistent with the
observed velocity width broadening at the filament's southern terminus, where
our line of sight will naturally intersect clouds that populate a broader
distribution of velocities, because some will be on their ascent, while others
will be slowing, and beginning to fall back inward.
That the filament's velocity is \textit{slower} near the nucleus than at
its high altitude terminus
suggests that gas has not fallen into it,
but rather has been lifted
out of it. For the two scenarios to be consistent with one another, then,
the filament should be dynamically young.
We will discuss the cavity uplift hypothesis in \autoref{sec:discussion}.

\begin{figure*}
 \begin{center}
  \includegraphics[scale=0.29]{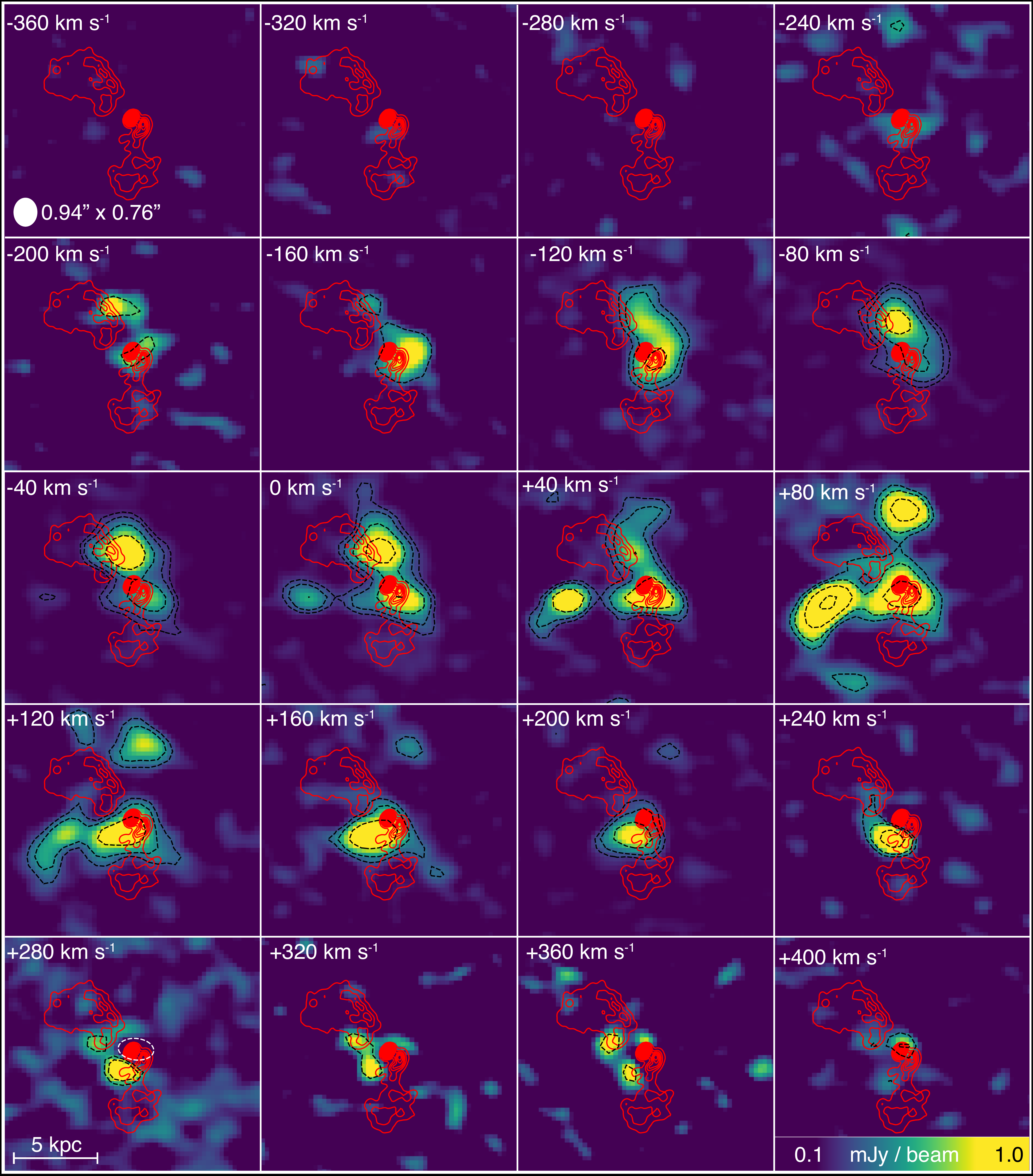}
 \end{center}
 \vspace*{-3mm}
 \caption{
  ALMA CO(2-1) channel maps, showing 40 \kms\ slices of the full data cube,
  ranging from $-360$ \kms\ through $+400$ \kms\ relative to the systemic
  velocity of the galaxy at $z=0.0821$. The outermost baselines have been tapered so as to increase
  signal to noise,
  resulting in beam size of $0\farcs94 \times 0\farcs79$,
  corresponding to a physical resolution of $1.4$ kpc $\times$ $1.2$ kpc
  (marked by the white ellipse in the top left panel).
  Red contours show the 8.4 GHz radio source, and dashed black contours
  are used to mark significance of the emission. The outermost black dashed
  contours show where the CO(2-1)
  emission exceeds $3\sigma$, and, when present, increase inward to show
  $5\sigma$, $10\sigma$, and $20\sigma$
  over the background RMS noise of 0.18 mJy beam$^{-1}$ per 40 \kms\ channel.
  The white dashed contour marks the continuum absorption discussed in \autoref{sec:natureresult}.
 }
 \label{fig:channel_maps}
\end{figure*}

\begin{figure*}
 \begin{center}
  \includegraphics[width=\textwidth]{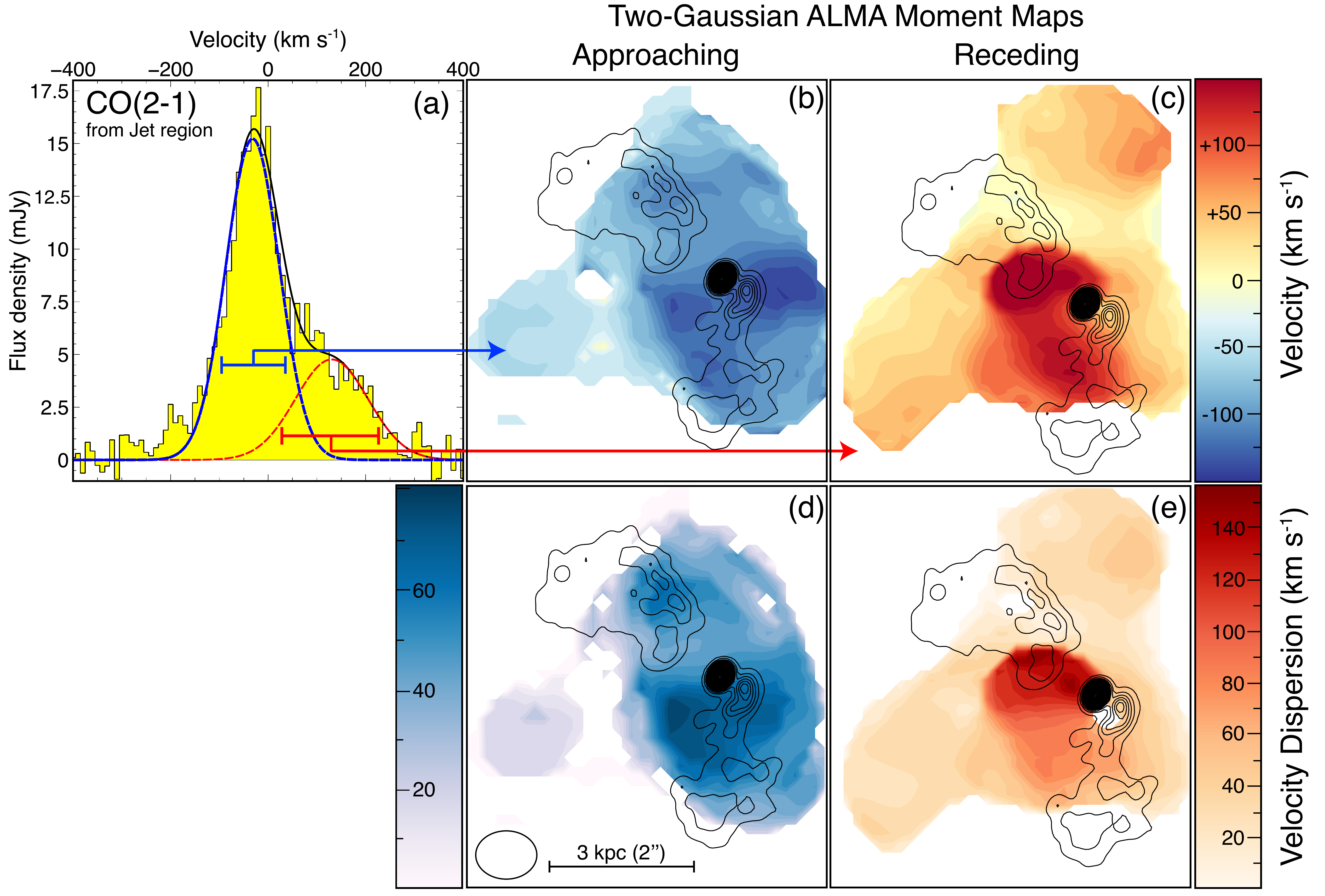}
  \vspace*{-9mm}
 \end{center}
 \caption{A closer look at the molecular gas cospatial with the radio jet. A single-Gaussian
  fit to this region (i.e., as shown for the moment maps in \autoref{fig:alma_momentmaps}) does
  not adequately model the superposition of approaching and receding components along the same line
  of sight. Here we show moment maps created with a double-Gaussian fit, better representing
  the velocity distribution.
  (\textit{a}) The CO(2-1) line
  profile extracted from a polygonal aperture encompassing the jet region, as shown in
  \autoref{fig:region_plots}\textit{b}. The line features a peak slightly blueward of center, as well as a strong red wing offset by $\sim+150$ km s\mone\ relative to the systemic velocity. Two Gaussians
  are fit to these components (shown in blue and red, respectively).
  Panels \textit{b} and \textit{c} show velocity maps for these approaching and receding
  molecular components, while panels \textit{d} and \textit{e} show their velocity
  dispersion maps.  Multi-Gaussian fits
  for various sub-regions are explored in \autoref{fig:jet_detail_expand}.  }
 \label{fig:fountain_overview}
\end{figure*}

\begin{figure*}
 \begin{center}
  \includegraphics[width=0.97\textwidth]{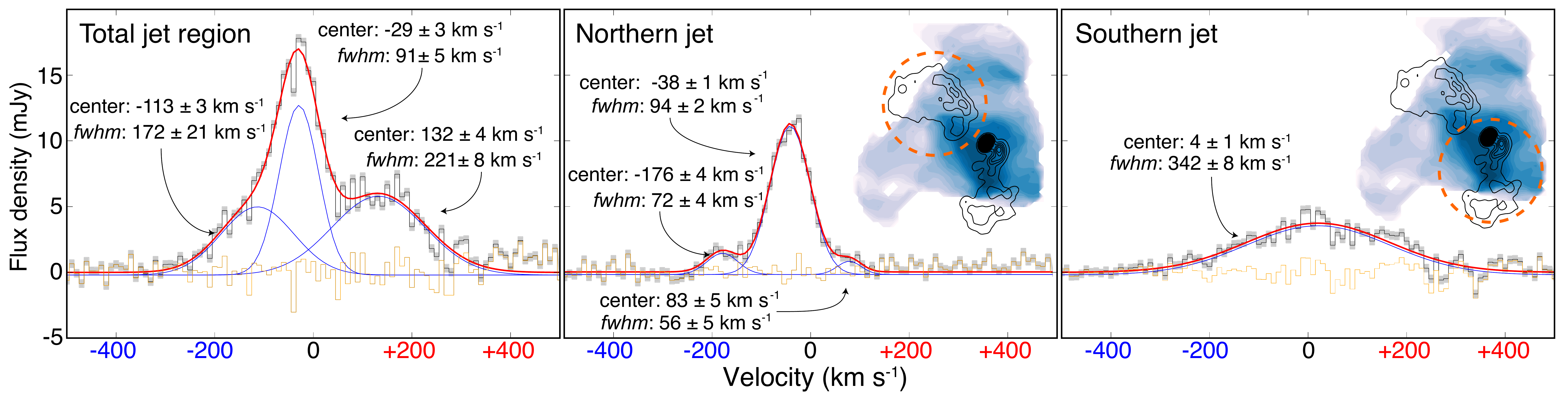}
  \vspace*{-6mm}
 \end{center}
 \caption{ALMA CO(2-1) spectra extracted from regions cospatial
 with the radio jet and lobes.
 One or more Gaussians have been fit to the data
 so as to minimize residuals, which are marked by the dark yellow line near 0 mJy.
 The (multi)-Gaussian fit is shown in red, while individual Gaussian components
 are shown in blue. The leftmost panel shows a three-Guassian fit to the entire
 region cospatial with the radio jet, while the center and righthand panels
 show fits to smaller regions cospatial with the northern and southern radio lobes,
 respectively. Those spectral extraction apertures are marked by orange circles
 on the in.aid velocity dispersion maps. Gaussian centroids and FWHMs for each component are labeled for all fits.}
 \label{fig:jet_detail_expand}
\end{figure*}

\begin{figure*}
 \begin{center}
  \includegraphics[width=\textwidth]{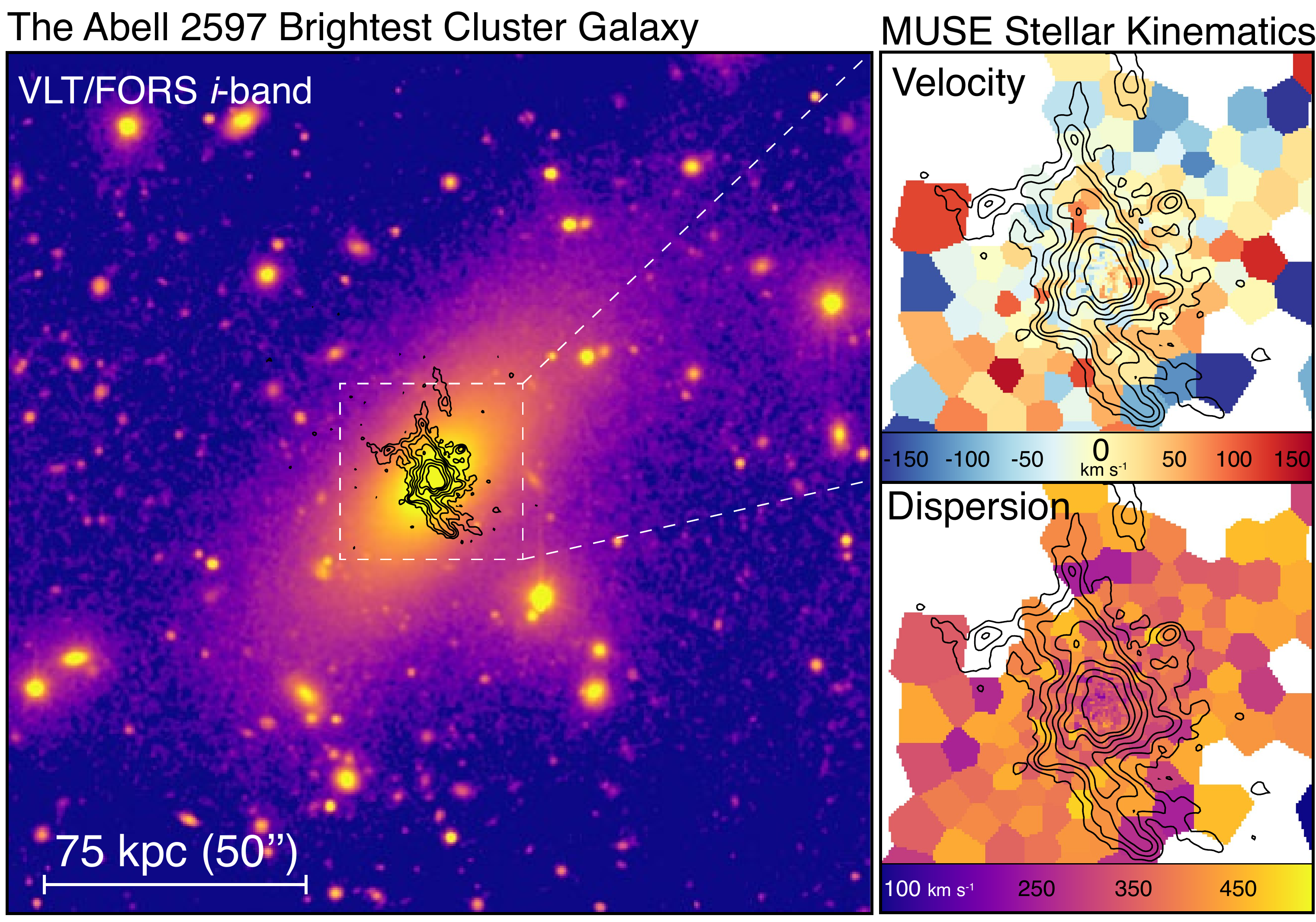}
 \end{center}
 \vspace*{-3mm}
 \caption{The host galaxy and the kinematics of its stellar component. (\textit{Left}) VLT/FORS $i$-band image of the BCG and its surrounding
  250 kpc $\times$ 250 kpc environment. A logarithmic stretch has been applied to
  highlight the low surface brightness outskirts of the galaxy.
  H$\alpha$ contours are shown in black, while the white
  dashed box indicates the FoV of the rightmost panels. (\textit{Top right})
  VLT/MUSE velocity map of the galaxy's stellar component.
  The data have been Voronoi binned so
  increase S/N in the stellar continuum, as described in \autoref{sec:musedata}.
  We only show the innermost
  $60\times60$ kpc$^2$ because the stellar surface brightness (and therefore S/N) drops rapidly beyond this FoV.
  Velocities have been projected around a zero-point at $z=0.0821$ (e.g., $cz=24,613$ km s\mone), as
  we have done for the ALMA and MUSE emission line velocity maps.
  (\textit{Bottom right}) Best-fit stellar velocity dispersion (e.g. FWHM$/2.35$), also from the MUSE data. Dispersions are typical for a large giant elliptical galaxy (e.g., \citealt{faber76}).  }
 \label{fig:MuseStars}
\end{figure*}

\begin{figure*}
 \begin{center}
  \includegraphics[width=\textwidth]{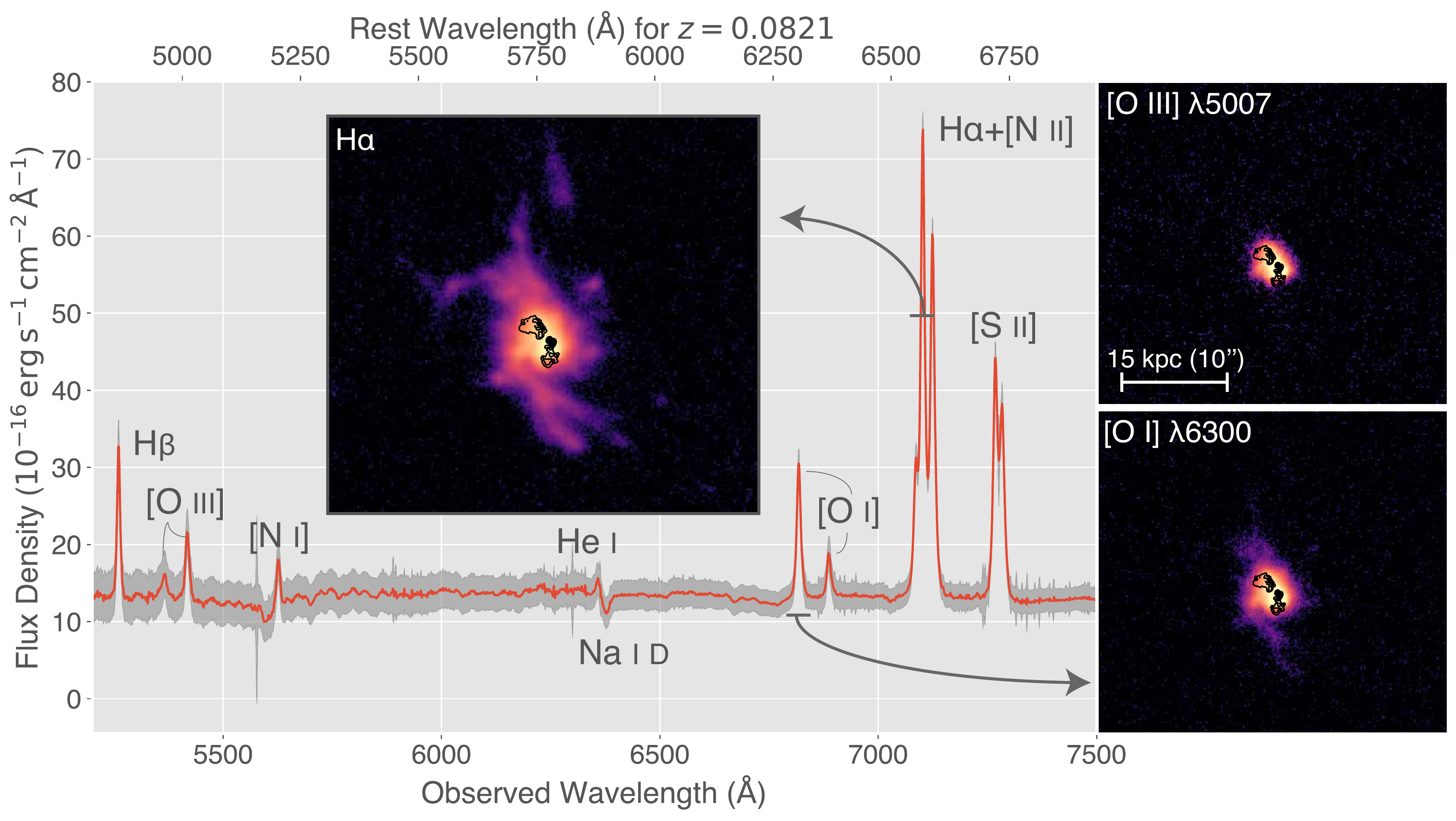}
 \end{center}
 \vspace*{-3mm}
 \caption{The MUSE optical spectrum extracted from a $10\arcsec$ circular aperture centered on the galaxy nucleus. Both nebular and stellar continuum emission are shown. The red-end of MUSE spectral coverage is around $9300$ \AA, but we have truncated it at $7500$ \AA\ for clarity. The MUSE IFU enables spatially resolved spectroscopy at the seeing limit ($\sim0\farcs9$) across the entire nebula, and so every spectral line here can be shown as a two dimensional image (or velocity/velocity dispersion map, e.g. \autoref{fig:MuseShowcase}). As examples, we show the continuum-subtracted H$\alpha$ image as an inset, as well as the [\ion{O}{3}]$\lambda5007$ and [\ion{O}{1}]$\lambda6300$ images to the right.   }
 \label{fig:MuseSpectrum}
\end{figure*}

\subsubsection{Cold gas motions induced by the radio jet}

The inner $\sim10$ kpc of the molecular nebula shows evidence for
dynamical interaction between the radio jet and the ambient molecular gas
through which it propagates.
This can be seen in \autoref{fig:channel_maps},
in which we show 40 km s\mone\ ``slices'' through
the CO(2-1) datacube (i.e., channel maps), from $-360$ km s\mone\ through
$+400$ km s\mone\ relative to the galaxy's systemic velocity.
The blueshifted channels reveal a sheet of
cold gas which, in projection, bends to hug the edges of the radio lobes
(see, e.g., the $-120$ km s\mone\ channel in \autoref{fig:channel_maps}, where the
alignment is most apparent).
The bulk
of this sheet's line-of-sight velocity is slow (only $\sim -100$ km s\mone),  though there is a thinner filament of higher velocity gas
that bisects the sheet lengthwise,  cospatial with a bright, linear knot along
a P.A. of $\sim 45^\circ$ (N through E) in the 8.4 GHz radio lobe.  The
velocity of this filament \textit{increases} (to $\gae 200$ km s\mone) with
increasing  galactocentric radius, which, like the southern filament (\autoref{sec:filament}),
is inconsistent with expectations of infall
under gravity.

The velocity structure of cold gas along the jet is better seen in \autoref{fig:fountain_overview}.
In panel \textit{a}, we show the CO(2-1) spectrum extracted from a $\sim10$ kpc (major axis)
elliptical aperture placed on the mm and radio core.
The line profile necessitates a fit with at least two Gaussians.
The emission associated with these two Gaussians is shown in panel \textit{b}.
A two-component velocity map, made by fitting the blue- and redshifted
components independently, is shown in panel \textit{c}.
The blueshifted shell of material, whose dispersion map
is shown in panel \textit{d}, is
bound on its northwestern edge by a linear ridge of higher velocity dispersion
blueshifted gas. In projection, this feature is cospatial  with the prominent
FUV-bright rim of star formation, detected by \textit{HST} (see \autoref{fig:overview}, bottom
right panel), that envelopes the northern radio lobe.  The molecular gas that is dynamically
interacting with the working surface of the radio jet is therefore likely
permeated by young stars.

As we noted in our discussion of \autoref{fig:pv_figure}\textit{b}, the broadest, fastest velocity structure
in the entire molecular nebula is cospatial with the bright radio knot at which
the southern radio jet bends sharply in position angle.
This is clearly evident in \autoref{fig:fountain_overview}, which shows
multi-Guassian fits to various spectral components of CO(2-1) emission
cospatial with the radio jet. These fits iteratively fit (and, if necessary, add)
Gaussians to the extracted spectra using a simple $\chi^2$ minimization technique.
The leftmost panel shows a three-Gaussian fit to the entire region cospatial with the
radio jet, while the center and right panels show fits to the regions
cospatial with the northern and southern radio lobes, respectively. The
spectral extraction apertures used are indicated by orange circles on the
images inlaid on these two panels.
A broad, single-Gaussian fit is needed for the region cospatial with the southern jet, including the location at which the jet is deflected.
This region includes the broadest velocity distribution of molecular gas in the galaxy,
with a FWHM of $342\pm8$ km s\mone ($\sigma=145\pm3$ km s\mone).
This fit has an integral of $\sim1.6\pm0.9$ Jy km s\mone, corresponding
to a molecular gas mass of  $\left(6.4\pm 0.4\right) \times 10^8$ \Msol.

\citet{pollack05} presented VLA polarimetry of
PKS $2322-123$, the radio source associated with the
A2597 BCG. The source has a steep spectral
index of $\alpha = 1.8$ between $\sim5$ and $\sim15$ GHz, suggesting
either that it is old or, given its compactness, that it has
remained dynamically confined as it struggles to expand against a dense, frustrating medium.
The VLA polarimetry reveals a compact region of
polarized flux associated with the southern lobe with a
Faraday rotation measure of 3620 rad m\mtwo,
suggesting that the southern lobe is deflected from its original
southwestern trajectory toward the south and into
our line of sight. This bright radio knot, cospatial with the broadest
velocity distribution of molecular gas (see \autoref{fig:fountain_overview} and \autoref{fig:jet_detail_expand}), is likely an impact site, showing strong evidence
for a dynamical interaction between the radio source and molecular gas. Whether
it is the molecular gas that has redirected the jet's trajectory
will be discussed in \autoref{sec:discussion}.

%

\begin{figure*}
 \begin{center}
  \includegraphics[width=\textwidth]{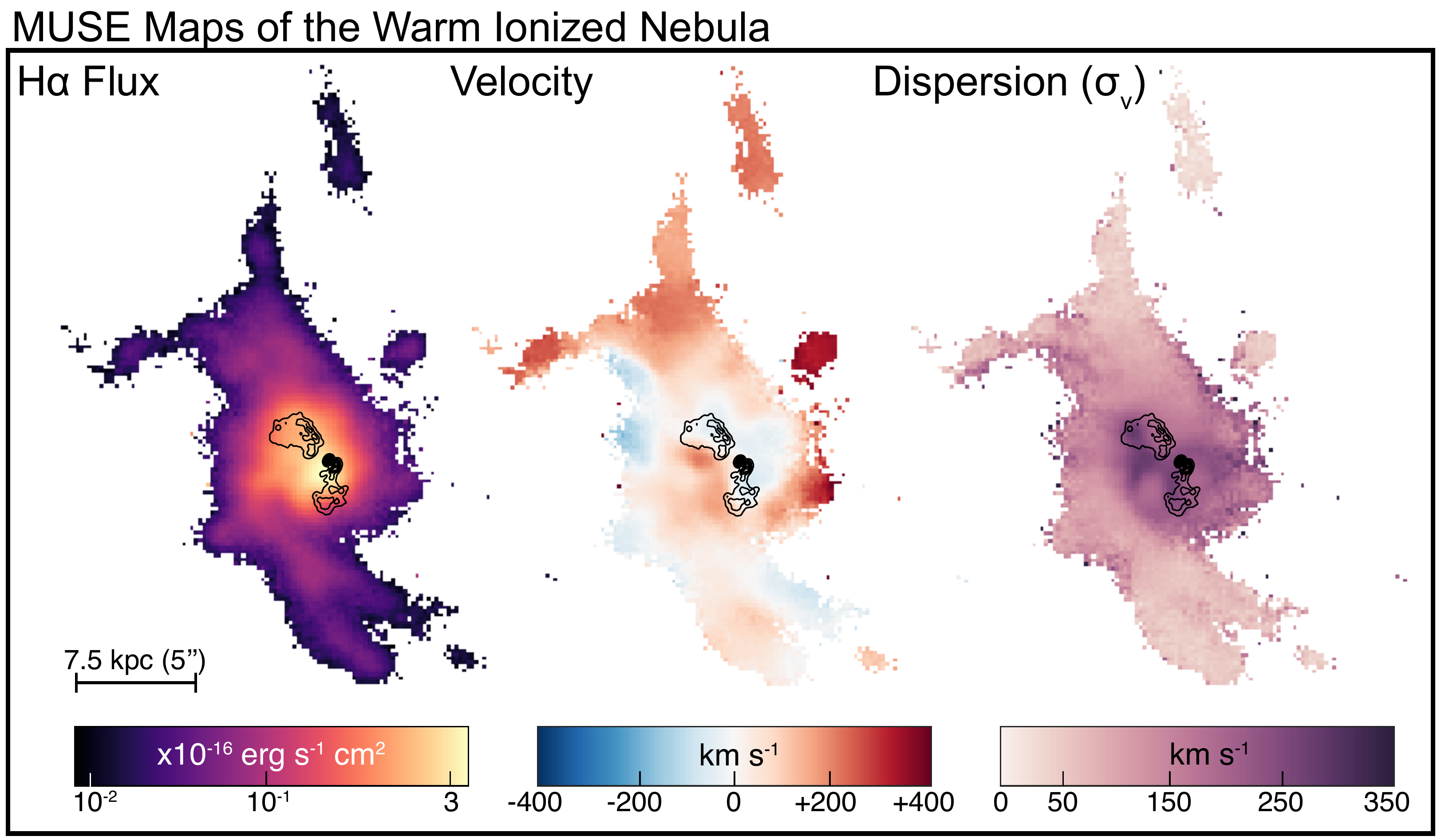}
 \end{center}
 \vspace*{-3mm}
 \caption{MUSE maps of H$\alpha$ flux, line of sight velocity, and velocity
  dispersion in the warm ionized nebula, created after modeling
  and subtracting the stellar continuum as described in \autoref{sec:musedata}.
  The H$\alpha$ flux map (left panel) is shown with a logarithmic color scale
  to better show the faint filaments relative to the bright nucleus.
  Note the
  blueshifted ``S''-shaped feature near the nucleus in the velocity map (center),
  strongly reminiscent of the shape of the 8.4 GHz radio source (shown in the flux map, for comparison). Note that these
  maps properly account for blending of the H$\alpha$ and [\ion{N}{2}] lines. Note also,
  particularly for the northern filaments, that velocities are higher at higher
  altitudes from the galaxy center, consistent more with uplift than gravitational freefall.
  Compare these maps to those for the cold molecular nebula in
  \autoref{fig:alma_momentmaps}. We compare the MUSE and ALMA data directly
  in \autoref{sec:musealma}.}
 \label{fig:MuseShowcase}
\end{figure*}

\begin{figure*}
 \begin{center}
  \includegraphics[width=0.9\textwidth]{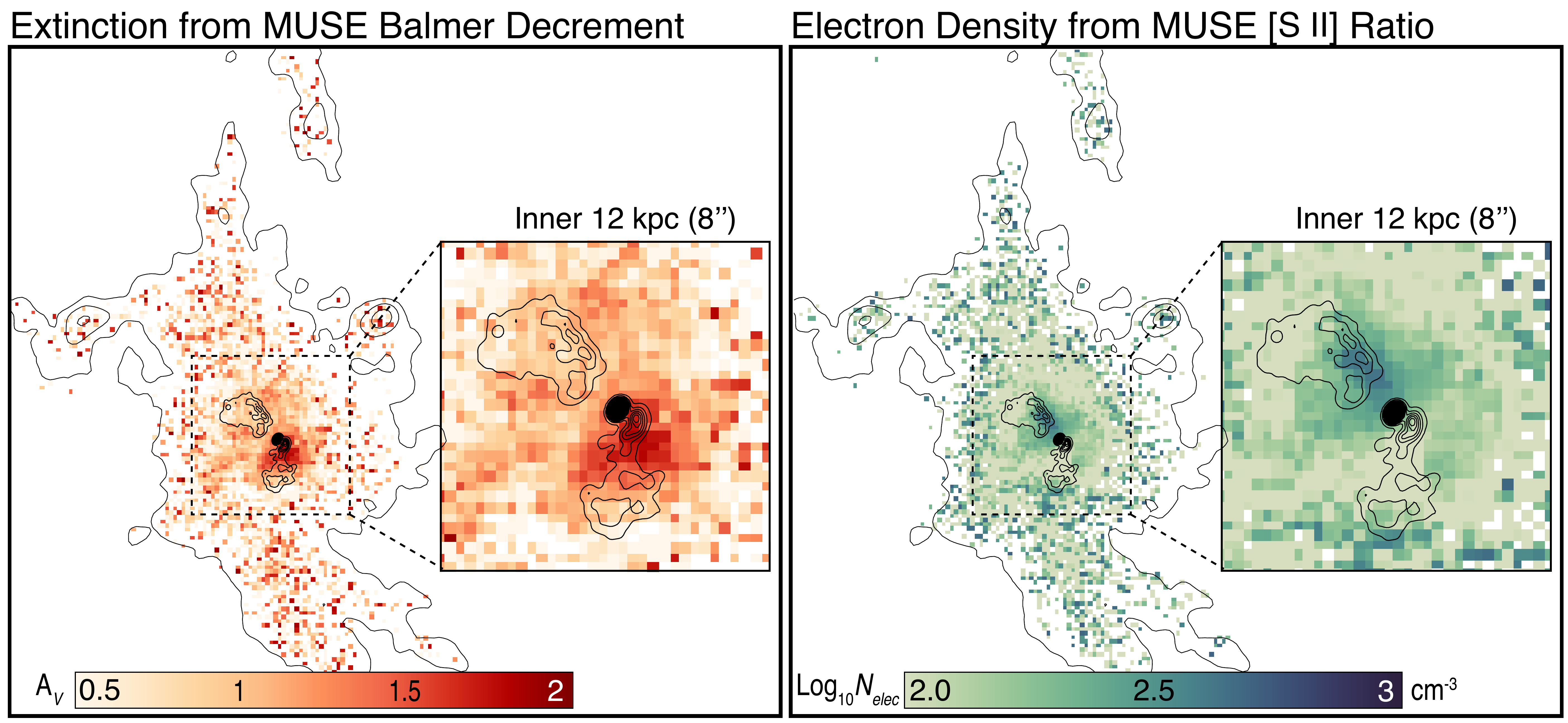}
 \end{center}
 \vspace*{-5mm}
 \caption{ (\textit{left}) An extinction ($A_V$)
  map made by scaling the MUSE Balmer decrement map (H$\alpha/$H$\beta$ ratio) following the procedure
  described at the end of \autoref{sec:musedata}. The 8.4 GHz radio source is overlaid in black contours. The inset panel shows a zoom-in on this $12\times12$ kpc$^2$ region. The color bar is in units of $V$-band magnitudes. (\textit{right}) Electron density map, made by scaling the ratio of the forbidden Sulfur lines
  ([\ion{S}{2}]$\lambda\lambda$ 6717 \AA\ / 6732 \AA) using the calibration of \citet{proxauf14} and
  assuming an electron temperature of $T_e=10^4$ K. The region of highest extinction is found
  just to the south of the nucleus, where the radio jet bends in position angle at the site of
  a bright radio knot with a large Faraday rotation measure, indicative of abrupt deflection. It is here that CO(2-1) is brightest (\autoref{fig:alma_momentmaps}).
  The electron density map is highest at the boundaries of the southern radio lobe,
  and along the long axis of the northern radio lobe. }
 \label{fig:extinction}
\end{figure*}

\begin{figure*}
 \begin{center}
  \includegraphics[width=0.9\textwidth]{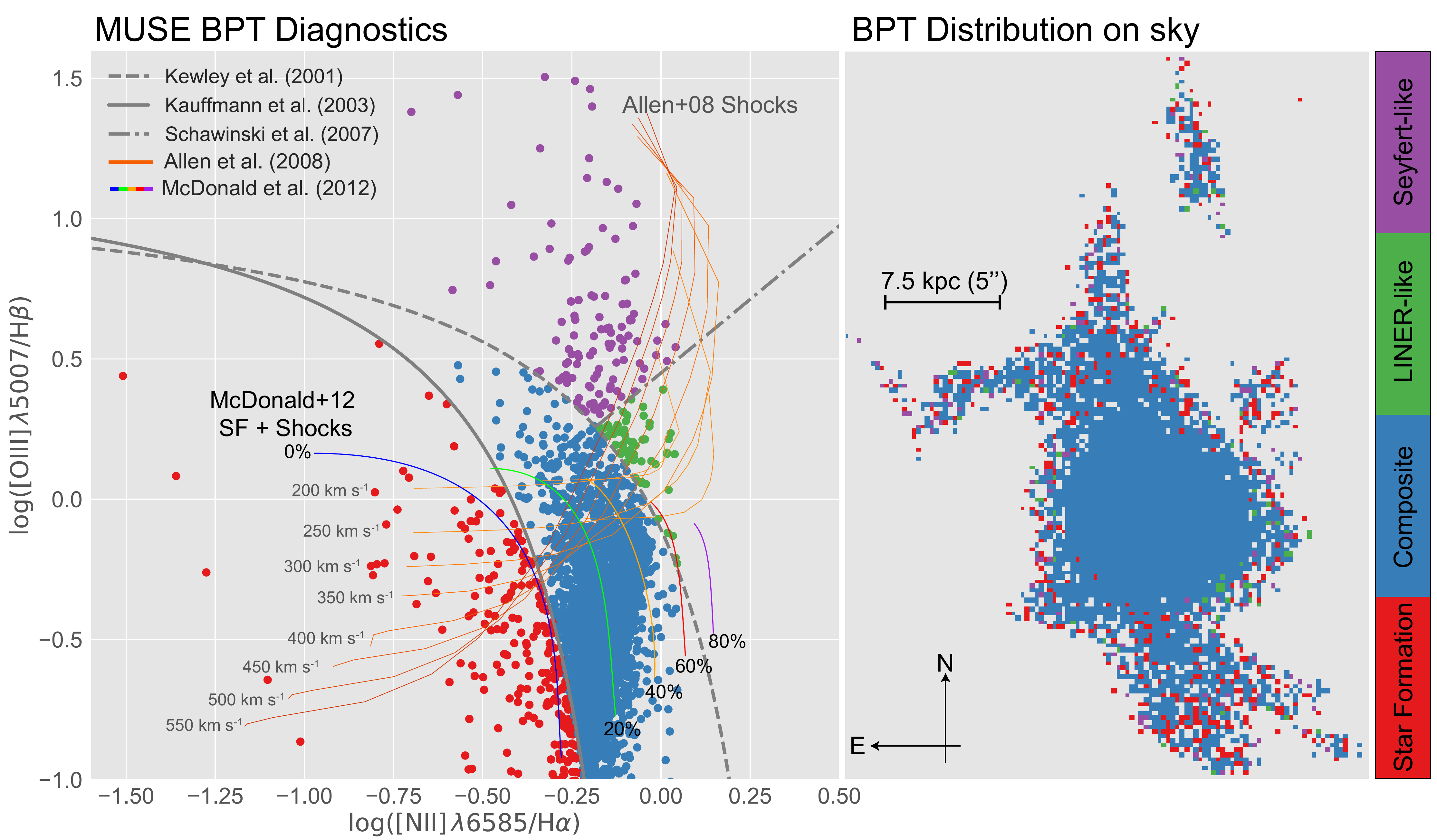}
 \end{center}
 \vspace*{-5mm}
 \caption{MUSE emission line diagnostic diagrams for spaxels with S/N$>3$ in each line.
 The left panel shows a standard Baldwin, Phillips, \& Terlevich (BPT; \citealt{baldwin81}) diagnostic plot using the [\ion{O}{3}]$\lambda5007$/H$\beta$ and [\ion{N}{2}]$\lambda6585$/H$\alpha$ line ratios (e.g., \citealt{veilleux87}).
 Spaxels are color-coded
 based upon their location relative to boundaries between well-known empirical and theoretical
 classification schemes \citep{kewley01,kauffmann03,schawinski07} shown in gray dashed and solid lines.
 We also show ``pure shock'' \citep{allen08} as well as ``slow shock + star formation'' \citep{mcdonald11b} composite models in solid color lines.
 We discuss these lines in \autoref{sec:muse}.
 Spaxel color coding is shown to the
 panel at right, which also shows their distribution on the sky. }
 \label{fig:bpt}
\end{figure*}

\begin{figure*}
 \begin{center}
  \includegraphics[width=0.98\textwidth]{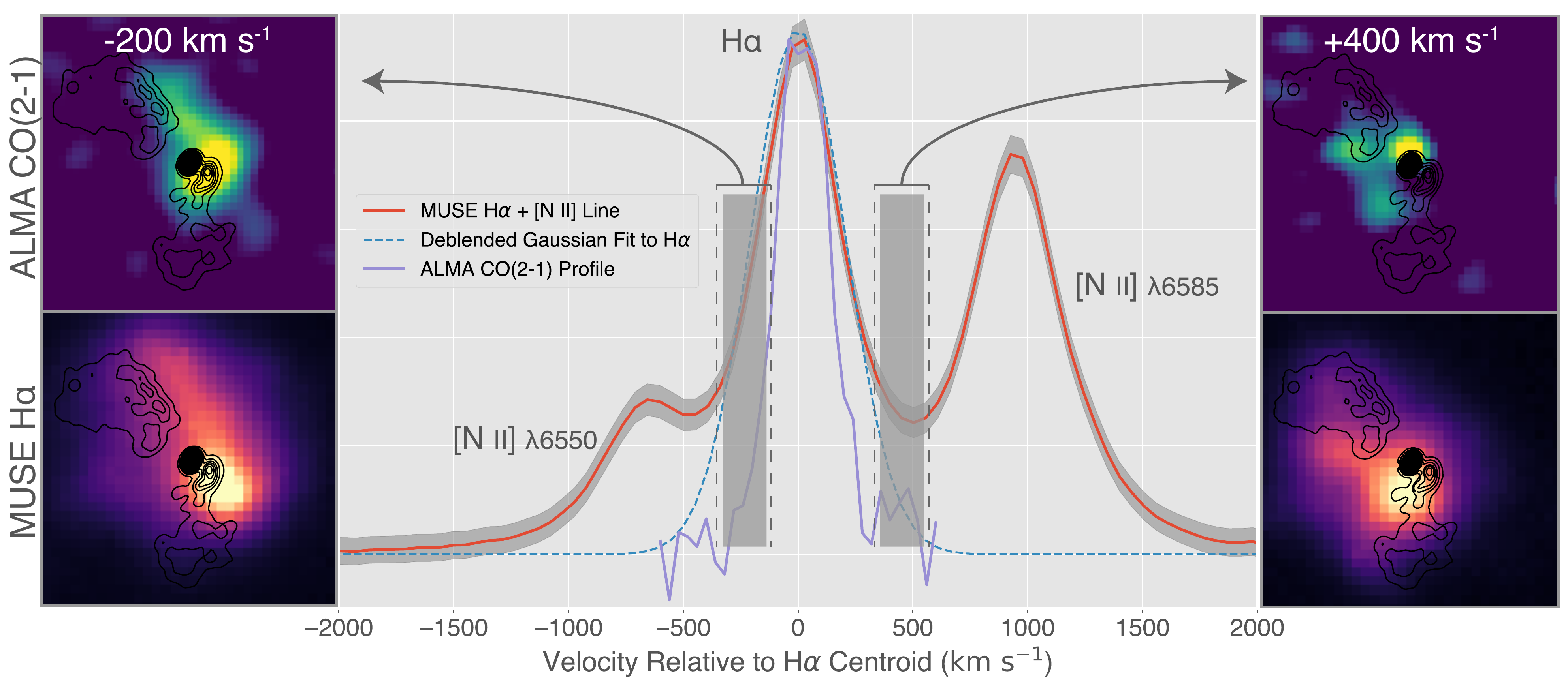}
 \end{center}
 \vspace*{-6mm}
 \caption{The MUSE H$\alpha$ and ALMA CO(2-1) datacubes reveal similar morphologies
 at matching velocities, consistent with the hypothesis that the
 warm ionized and cold molecular gas are co-moving with one another, as would be predicted
 if the H$\alpha$ emission arose from the warm ionized skins of mm-bright
 molecular cores. Here we show the MUSE H$\alpha$+[N~\textsc{ii}] profile
 extracted from a circular aperture with a diameter of 30 spaxels ($\sim6\arcsec$), centered
 on the galaxy core. We have deblended the H$\alpha$ line from the [N~\textsc{ii}] doublet,
 and plot the resulting single Gaussian fit to H$\alpha$ with the blue dashed line.
 The ALMA CO(2-1) spectrum, extracted from a (roughly) matching aperture,
 is plotted in purple.}
 \label{fig:MuseALMA}
\end{figure*}

\begin{figure*}
 \begin{center}
  \includegraphics[width=0.98\textwidth]{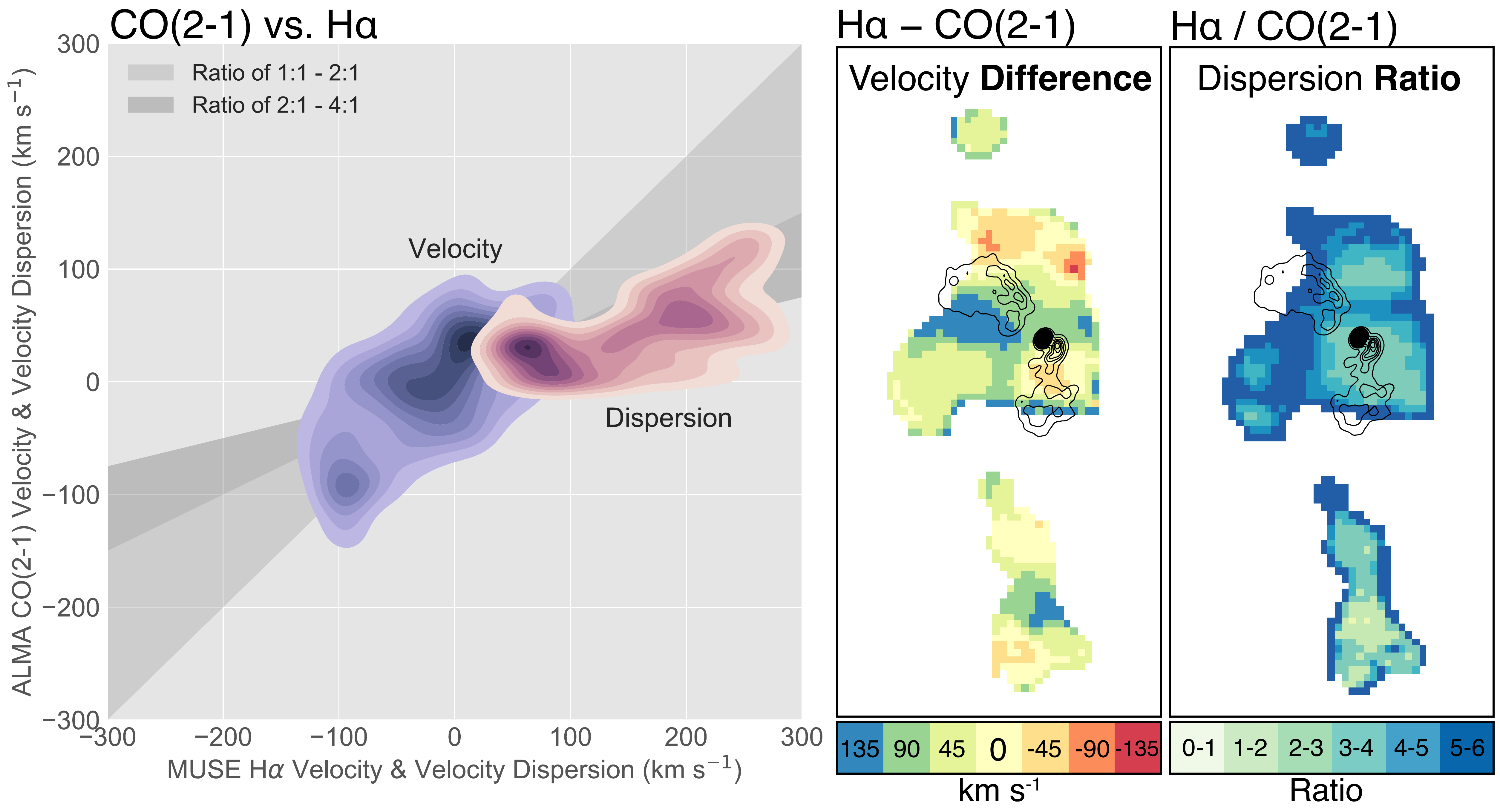}
 \end{center}
 \vspace*{-3mm}
 \caption{
 (\textit{Left panel}) ALMA CO(2-1) vs.~MUSE H$\alpha$ velocity and velocity dispersion. Points
  are taken from every cospatial spaxel in the $>3\sigma$ overlap region between the
  warm ionized and cold molecular nebulae. The points have been smoothed with a Gaussian
  kernel. Contour colors encode the Gaussian kernel density estimate (i.e., a darker color indicates a higher density of data points).
 (\textit{Right panels})
 Maps of the difference and ratio between H$\alpha$ and CO(2-1) velocity centroid and
  dispersion, made by
  subtracting and dividing the corresponding MUSE and ALMA moment maps, respectively. We have applied
  various corrections to account for, e.g., differing spatial resolutions,
  as described in \autoref{sec:ratiomaps}. The
 edges of these maps should be ignored. In the dispersion ratio map, for example,
 the outermost dark blue rim is smaller than the ALMA beam size, and an artifact of the division.   }
 \label{fig:MuseALMARatios}
\end{figure*}

\subsection{MUSE maps of the host galaxy and warm nebula} \label{sec:muse}

In \autoref{fig:MuseStars} we show the Voronoi-binned MUSE map of stellar line
of sight velocity and velocity dispersion within the inner 50 kpc of the galaxy.
Only Voronoi-binned spaxels with S/N$>200$ are shown.
A deep VLT/FORS $i$-band image of the BCG with
MUSE H$\alpha$ contours overlaid is shown for reference. While some
background/foreground sources are seen, there are a number of spectroscopically confirmed companions
embedded within the stellar envelope of the BCG (Tremblay et al.~in prep.). The galaxy has clearly enjoyed a rich
merger history, as is generally the case for all those that sit long enough at the bottom of a cluster potential well.
At best, there is only a weak signature of coherent stellar rotation in the inner 50 kpc (NW approaching, SE receding),
consistent with expectations for
the boxy interior of a ``slow / non-regular rotator'' early type galaxy \citep{cappellari16}.
We note that there is some evidence for a minor-axis kinematically
decoupled core (KDC, e.g., \citealt{kranovic11}) in the nucleus. This will
be discussed in a forthcoming paper (Tremblay et al.~in prep.).

The total (stellar and nebular) spectrum, extracted from a spatial aperture that
encompasses the galaxy center in the MUSE cube, is shown in
\autoref{fig:MuseSpectrum}. All major nebular lines are detected
at high signal-to-noise, enabling spatially resolved line maps from H$\beta$ through [\ion{S}{2}].
A selection of these are shown in the side panels of \autoref{fig:MuseSpectrum}.
We note that [\ion{O}{3}]$\lambda\lambda4959,5007$ \AA\ is spatially extended,
but only on the scale of the 10 kpc 8.4 GHz radio source. The remaining lines,
particularly those tracing star formation, match the morphology (and linewidth, roughly) of the H$\alpha$
nebula, albeit at lower surface brightness.
As is apparent from \autoref{fig:MuseStars}, the major axis of the warm emission line nebula is roughly aligned
with the stellar minor axis of the host galaxy.

In \autoref{fig:MuseShowcase} we show the MUSE flux, velocity, and velocity
dispersion maps for the H$\alpha$ nebula.
Just as for the cold molecular gas, the warm ionized nebula has not dynamically equilibrated, as there are no obvious signs of
rotation save for the innermost $\sim10$ kpc of the galaxy.
There, a blueshifted shell of material is found, cospatial with a similar
feature in the molecular gas, clearly matching the shape
of the 8.4 GHz radio source. The H$\alpha$ velocity dispersion map reveals thin,
bubble-like rims of higher velocity dispersion gas, reaching widths upward
of  $\sim350$ km s\mone. Given their location and morphology,
these broad streams are likely churned by dynamical interaction with the radio source,
or the buoyant X-ray cavities it has inflated.
Cospatial with these features, \citet{oonk10} discovered coherent velocity streams of warm molecular hydrogen (traced by the H$_2$ 1-0 S(3) and Pa$\alpha$ lines) similarly
hugging the edges of the radio source, at roughly
the same line of sight velocity and velocity width as those seen in the MUSE H$\alpha$ maps.


Again, like the molecular nebula, the northern and southern
warm ionized filaments are more difficult to interpret.
All show narrow velocity structure, and no evidence for freefall.
This is the case for a large number of warm nebulae in CC BCGs \citep{hatch07,edwards09,hamer16},
even those for which there is extremely compelling morphological coincidence
between filaments and X-ray cavities, suggestive of uplift (see, e.g.,
IFU observations of the ionized filaments in Perseus, \citealt{hatch06, gendron-marsolais18}).
As we will discuss in \autoref{sec:discussion}, A2597 is in many ways like Perseus in
that the H$\alpha$ filaments are spatially
coincident with X-ray cavities.
We discuss these implications in \autoref{sec:discussion}.

Dividing the MUSE H$\alpha$ and H$\beta$ flux density maps produces a Balmer Decrement map which, following the assumptions discussed in \autoref{sec:musedata}, we scale to create the extinction ($A_V$) map shown in \autoref{fig:extinction}\textit{a}. The highest extinction, and therefore perhaps
the densest, dustiest gas, is found to the south of the nucleus, where the radio
source is deflected. CO(2-1) is brightest at this same knot (compare the ALMA moment zero map in \autoref{fig:alma_momentmaps} with \autoref{fig:extinction}). The northeastern dust lane seen in optical imaging
is also clear. It is along this rim that we find extended 230 GHz continuum emission
(see \autoref{fig:continuum}). This could indeed be dust continuum emission,
detected at $\sim3\sigma$ alongside the $\sim425\sigma$ non-thermal mm-synchrotron
point source associated with the AGN. \autoref{fig:extinction}\textit{b} shows
the electron density map (linearly proportional to the total gas density at $\sim10^4$ K) made from the [\ion{S}{2}]$\lambda\lambda$ 6717 \AA\ / 6732 \AA\ line ratio.  The densest
gas is found along the jet axis, perhaps due to dredge-up of cooler, more dense ionized gas from the nucleus, and also along
a southerly ``shell'' that appears to hug the boundary
of the southern radio jet as it bends in position angle. If real (and it likely is, given that it is also seen in the $A_V$ map made from the Balmer lines), this may be
tracing the dense population of clouds that form the impact site at which
the jet is deflected. We will discuss this possibility in \autoref{sec:discussion}.

In \autoref{fig:bpt} we show a spatially resolved Baldwin, Phillips, \& Terlevich (BPT; \citealt{baldwin81}) diagnostic plot using the [\ion{O}{3}]$\lambda5007$/H$\beta$ and [\ion{N}{2}]$\lambda6585$/H$\alpha$ line ratios  (e.g., \citealt{veilleux87}) extracted from each spaxel in the MUSE cube.
Galaxies (or indiviual regions within a single galaxy, as shown here) stratify in BPT space
based upon the relative contributions of stellar and non-stellar ionization sources.
The solid gray curve shows the empirical star formation line from \citealt{kauffmann03}, the dashed gray curve shows the theoretical maximum starburst model of \citealt{kewley01}, and the dash-dotted gray line is the empirical division between LINER- and Seyfert-like sources as defined by \citealt{schawinski07}.
We have color-coded the data points based upon the regions in which they sit. The vast majority of points
lie in the ``composite'' or ``AGN-H\textsc{ii}'' region as defined by \citealt{kewley06} (also called the ``transition region'' in \citealt{schawinski07}). This ``classification'' should not be over-interpreted, as the situation for CC BCGs is highly complex and likely represents a superposition of several different ionization sources (see, e.g., the discussions of \citealt{ferland09,mcdonald11b}. To illustrate this,  we plot lines of constant
shock velocity (in orange) from the \citet{allen08} library of fast radiative shocks, assuming a gas density of $n=1000$ cm$^{-3}$ (i.e., roughly the value in the central regions of \autoref{fig:extinction}\textit{b}), as well as the ``slow shock + star formation'' composite models adapted from \citet{farage10,mcdonald11b}.
Debate continues as to the relative role played by stellar photoionization,
(slow) shocks \citep{mcdonald11b}, conduction \citep{sparks12}, and cosmic ray heating \citep{ferland09,donahue11,fabian11a,mittal11,johnstone12}.
Galaxies are enormous,  complex structures, and so any line of sight
that passes through them will inevitably reveal a superposition of many
physical processes. It is likely that all of these ionization mechanisms
play some role in heating the envelopes of cold clouds. We note, finally,
that new \textit{HST}/COS far-ultraviolet spectroscopy of the filaments in A2597
will be discussed in a forthcoming paper (Vaddi et al.~in prep.).

\subsection{MUSE and ALMA Comparison} \label{sec:musealma}

Comparing the MUSE and ALMA data directly reveals strong evidence
that the warm ionized and cold molecular nebulae are not only cospatial,
they are comoving.
In \autoref{fig:MuseALMA} we overplot the H$\alpha$+[\ion{N}{2}]
and CO(2-1) profiles extracted from matching 6\arcsec\ diameter
apertures centered on the galaxy core in the MUSE and ALMA cubes,
respectively.
The panels at the sides of \autoref{fig:MuseALMA} show matching H$\alpha$ and CO(2-1)
morphology at the broadest wings of each line, consistent (but not \textit{uniquely}) consistent
with the hypothesis that the two lines stem from largely the same population of clouds.
This cannot be true entirely, as the deblended H$\alpha$ FWHM is $565 \pm 25$ km s\mone, a factor of $\sim2$ broader
than the $252\pm14$ km s\mone\ FWHM of the CO(2-1) line.

This velocity width mismatch is more readily apparent in \autoref{fig:MuseALMARatios},
where we plot CO(2-1) line-of-sight velocity and dispersion against the same
quantities for H$\alpha$. We have smoothed the data points (i.e., one point for each cospatial
spaxel in the registered MUSE and ALMA cubes) with a Gaussian, and show
shaded regions indicating ratios of  1:1-2:1 and 2:1-4:1. While the line velocity centroids
lie largely along the 1:1 line, the line widths preferentially span the 2:1-4:1 range.
The rightmost panels of \autoref{fig:MuseALMARatios} show the difference and ratio, respectively, between the MUSE H$\alpha$ and ALMA CO(2-1)
velocity and dispersion maps.
In the velocity difference map, bluer colors
mean that the CO(2-1) velocity centroid is slightly blueshifted relative to the H$\alpha$ velocity centroid. The velocity
difference map is largely smooth and below $\pm45$ km s\mone, which shows that the H$\alpha$ and CO(2-1) line velocity centroids track one another closely across the entire overlap region between the molecular and ionized nebulae. The velocity
dispersion ratio map (\autoref{fig:MuseALMARatios}, right panel) shows that, on average, the H$\alpha$ velocity
dispersion is a factor of $2-3$ times broader than that for CO(2-1).

The broader observed velocity widths for H$\alpha$ are important but
not \textit{necessarily} surprising,  given that
our line of sight is likely to intersect more warm gas (and therefore a broader velocity distribution) than it is for the cold molecular clouds,
owing to their large relative contrast in volume filling factor.
We discuss this further in \autoref{sec:discussion}.

\section{Discussion}

\label{sec:discussion}

This paper presents three results:

\begin{enumerate}

 \item \textbf{Cold gas is cospatial and comoving with warm gas}. A three billion solar mass filamentary molecular nebula is found to span
       the inner 30 kpc of the galaxy. Limited by the critical density of CO(2-1), its volume filling factor must be low, and so the nebula must be more like a ``mist'' than a monolithic slab of cold gas (e.g., \citealt{mccourt18}). These cold clouds are likely wrapped in warm envelopes that shine with Balmer and forbidden line emission at the cloud's interface with the hot X-ray atmosphere,
       explaining why the H$\alpha$ and CO(2-1) nebulae are largely cospatial and comoving. This hypothesis is now supported by a large and
       growing number of ALMA observations of CC BCGs (e.g., papers by Russell, McNamara and collaborators).

 \item \textbf{Cold gas is moving inward, and perhaps feeding the black hole}. Clouds are directly observed to fall inward toward the galaxy nucleus, probably within close proximity ($\lae 100$ pc)
       to the central supermassive black hole. These clouds may therefore provide a substantial (even dominant) component of the mass flux toward the black hole accretion reservoir. This result, discussed in \citealt{tremblay16} and considered in a broader context here, is consistent with a major prediction of the chaotic cold accretion (CCA) model \citep{gaspari13}.

 \item \textbf{Cold gas is dynamically coupled to mechanical black hole feedback}. In projection, a bright rim of blueshifted molecular gas
       appears to encase the radio lobes  (see e.g., \autoref{fig:channel_maps}), perhaps suggestive of dynamical
       coupling between the cold molecular gas and the powerful radio jet plowing through it.
       The broadest distribution of cold gas velocities is found copatial with the southern jet (\autoref{fig:jet_detail_expand}, right panel). Just south of the radio core, this jet deflects in position angle, perhaps because it has exchanged momentum with a dense ensemble of cold clouds. Nearly all cloud velocities, save for the most extreme wings
       of the distribution, are nevertheless below the circular speed at any given radius, and so the clouds
       should be falling inward unless tethered to the hot medium. $\sim1$ billion \Msol\ of cold gas is found in dynamically short-lived filaments spanning altitudes greater than 10 kpc from the galaxy center, and may be draped around the rims of buoyant X-ray cavities. We argue that effectively all
       of these non-equilibrium cold gas structures are directly or indirectly due to mechanical black hole feedback, as mediated either
       by jets, buoyant hot cavities, or turbulence in the velocity field of the hot atmosphere.

\end{enumerate}

It is possible that the molecular and ionized nebula at the heart of Abell 2597 is effectively a galaxy-scale ``fountain'', wherein cold gas drains into the black hole accretion reservoir,
powering a jet- or cavity-driven plume of uplifted low-entropy gas that ultimately
rains back toward the galaxy center from which it came.
This scenario might establish
a long-lived heating-cooling feedback loop, mediated by the supermassive black hole, which would act
much like a mechanical ``pump'' for this fountain.

\subsection{The Fountain's ``Drain''}
%

We directly observe at least three cold molecular clouds moving toward what would
be the fountain's drain (see \autoref{sec:natureresult} and \citealt{tremblay16}).
If this line-of-sight observation is at all representative
of a (much) larger three-dimensional distribution of inward-moving clouds, and if indeed
they are as close to the black hole as corroborating evidence suggests they are,
they could supply on the order of $\sim0.1$ to a few \Msol\ yr\mone\
of cold gas to the black hole's fuel reservoir.
The observation would then be consistent with a major prediction of
\citet{gaspari13,gaspari15,gaspari17b}, who argue that
that nonlinear condensation from a turbulent,
stratified hot halo induces a cascade of multiphase gas that condenses from the $\sim10^7$ K to the $\sim20$ K regime.
This cooling ``rain'' manifests as chaotic motions that dominate over coherent rotation
(with turbulent Taylor number $<1$; e.g., \citealt{gaspari15}).
Warm filaments condense along large-scale turbulent eddies
(generated, for example, by AGN feedback), naturally creating extended and
elongated structures like the H$\alpha$ filaments ubiquitously observed in CC BCGs, and possibly
explaining their apparent close spatial association with radio jets and X-ray cavities (e.g., \citealt{tremblay15}).
Warm overdensity peaks further condense into many cold molecular clouds\footnote{Though the need for dust grains to act as a catalyst for the formation of molecular gas remains a persistent issue, e.g., \citealt{fabian94,voit11}.}, hosting
most of the total mass, that form giant associations.
The thermodynamics and kinematics of the cooler gas phases should then retain
``memory'' of the hot plasma from which they have condensed \citep{gaspari17b,gaspari18,voit18b}.

Despite important differences (reviewed in part by \citealt{hogan17,gaspari18,voit18b,pulido18}),
the chaotic cold accretion model of \citealt{gaspari13} succeeds alongside the ``circumgalactic
precipitation'' and ``stimulated feedback'' models of \citet{voit15b} and \citet{mcnamara16} (respectively) in predicting many of the major observational results we find in Abell 2597.
Were we to (roughly) attempt to unify these models within the same ``fountain'' analogy,
all would effectively include a ``drain'' into which cold clouds fall, providing
a substantial (even dominant) mass flux toward the black hole fuel reservoir. That we have strong observational evidence for exactly
such a drain in Abell 2597 enables us to place at least broad constraints on how the drain might operate.

For example, whether they condense in the turbulent eddies of cavity wakes or not,
a cascade of gas cooling from  hot plasma will still require roughly a cooling time $t_\mathrm{cool}$ to reach the molecular phase.
Using the buoyant rise time as a rough age estimate, the oldest X-ray cavities in A2597 are $\sim2\times10^8$ yr \citep{tremblay12b}, which is roughly
comparable to the cooling time at the same $20-30$ kpc radius \citep{tremblay12a}.
The time it takes for clouds to descend from any given altitude to the center of the galaxy is a
more complicated issue.
Following \citealt{lim08}, a thermal instability, precipitating at rest with respect to the local ICM velocity,
will freefall in response to the gravitational potential and accelerate to a velocity $v$
given roughly by
\begin{equation}
 v = \sqrt{v(r_0)^2 + 2GM \left( \frac{1}{r + a} - \frac{1}{r_0 - a} \right)},
\end{equation}
where $v(r_0)$ is its initial velocity (assumed to be zero if the ICM and BCG velocities are roughly matched), $r_0$ is its
starting radius relative to the BCG core, $G$ is the gravitational constant, $M$ is the total gravitating mass of the BCG, and $a$
is its scale radius (which is roughly half the effective radius $R_e$, as $a\approx R_e / 1.815$).
For a scale radius of $a\sim20$ kpc and a gravitating mass of $M\approx10^{12}$ \Msol\ \citep{tremblay12a},
the cooling cloud would attain a rough velocity of $\sim470$ km s\mone, $\sim380$ km s\mone, or $\sim300$ km s\mone\ if
it fell from a height of 20, 10 or 5 kpc, respectively.

Observed line-of-sight cloud velocities in Abell 2597 are significantly
lower than these freefall values, just as they are for effectively all other CC BCGs thus far
observed with ALMA (see e.g. \citealt{vantyghem18}, for the latest example).
The clouds might still be ballistic if most of their motion is contained in the plane of the sky,
but this argument weakens with every new observation showing the same result.
It is therefore now clear that the velocity of cold clouds in the hot atmospheres of CC BCGs
cannot be governed by gravity alone.  Simulations and arguments by (e.g.) \citet{gaspari18} and \citet{li18} indeed suggest that the clouds must have sub-virial velocities, consistent with
those observed in CC BCGs including Abell 2597 (\autoref{sec:velocitystructure}).

If a cooling cloud's terminal speed is smaller than typical infall speeds \citep{mcnamara16}, it can
drift in the macro-scale turbulent velocity field of the hot X-ray atmosphere \citep{gaspari18},
whose dynamical structure is sculpted by jets, sound waves, and bubbles.
The terminal velocity of cold clouds is set by the balance of their weight
against the ram pressure of the medium through which they move (e.g., \citealt{li18}).
That the clouds in Abell 2597 are apparently not in freefall may simply mean
that their terminal velocity is the lower of the two speeds.
While the extreme density contrast between molecular gas and hot plasma remains
an issue, one simple explanation is that the clouds' velocity in the hot atmosphere
has been arrested by more efficient coupling mediated by their warm ionized
skins, which would effectively lower their average density (and therefore their terminal speed) and increase
the strength of any magnetic interaction (e.g., \citealt{fabian08}).

Given the apparent lack of coherent velocity gradients along the molecular
and ionized filaments, it is also likely that the multiphase nebula is dynamically
young. Such a result is unsurprising in the context of chaotic cold accretion,
precipitation, and stimulated feedback models. In essence, all suggest that
the cold clouds are just one manifestation of what is ultimately
the same hydrodynamical flow, drifting in the velocity field of the hot plasma.
That velocity field, in turn, is continually stirred by
subsonic turbulence induced by
buoyant bubbles, jets, and merger-driven sloshing \citep{gaspari18}.
This omni-present dynamical mixing may inhibit virialization, preventing
the formation of smooth gradients over kpc scales.
At the very least, the recent \textit{Hitomi} observation of Perseus confirms
that bulk shear in the hot plasma is similar to molecular gas speeds observed
with ALMA,
supporting the idea that they move together \citep{hitomi16}.
At sub-kpc scales, inelastic collisions and tidal stress
between clouds can funnel cold gas toward the nucleus,
which we observe directly in Abell 2597 (\autoref{sec:natureresult} and \citealt{tremblay16}).  Chaotic cold accretion can then boost
black hole feeding far in excess of the Bondi rate, powering
the ``pump'' at the fountain's center.

\begin{figure*}
 \begin{center}
  \includegraphics[scale=0.17]{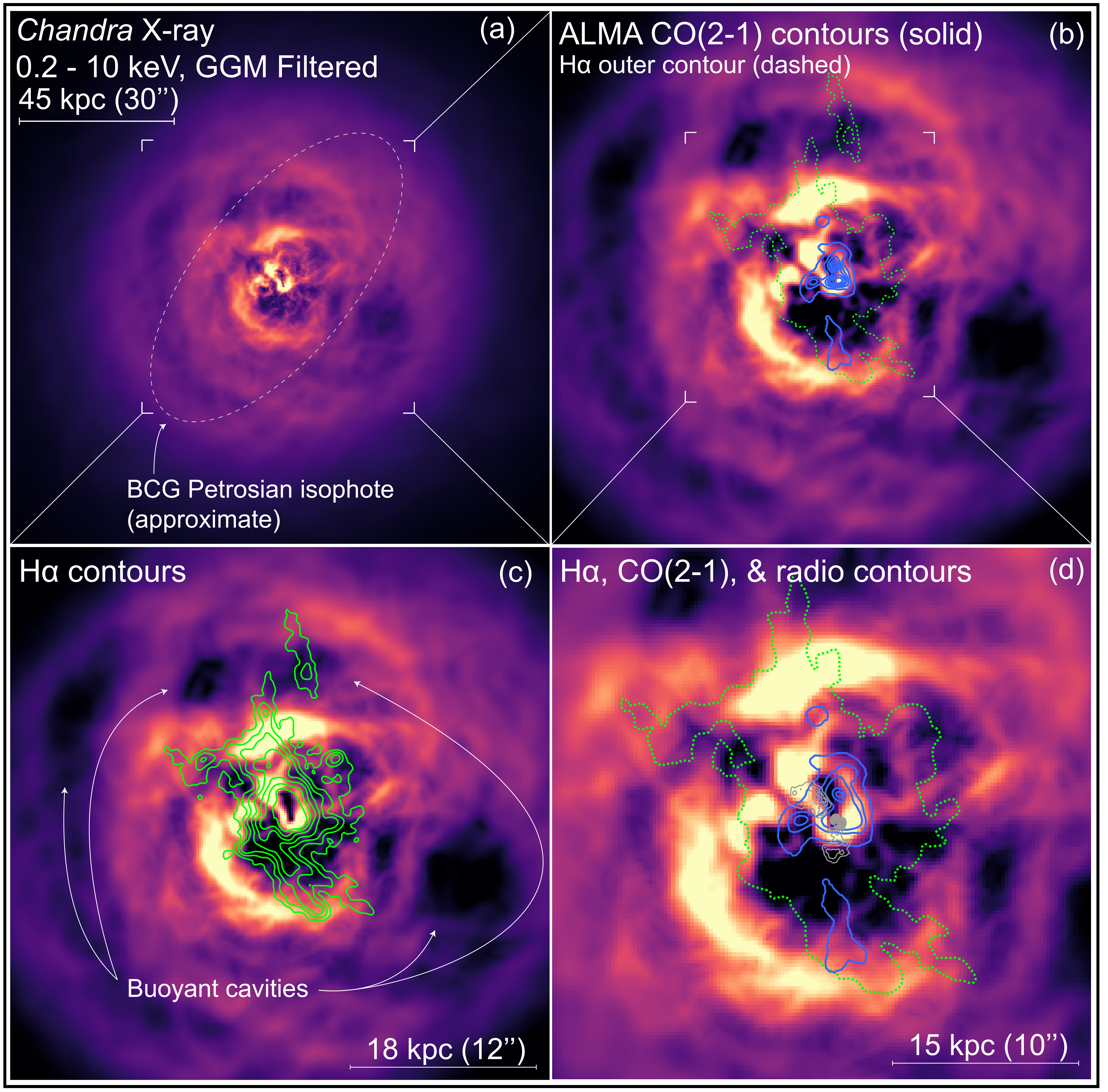}
 \end{center}
 \vspace*{-4mm}
 \caption{
  A new, deeper look at the X-ray cool core cospatial with the A2597 BCG. (\textit{a}) 626 ksec \textit{Chandra} X-ray observation of $0.2 - 10$ keV
  emission in the innermost $\sim250\times250$ kpc$^2$ of the cluster.  The X-ray data have been
  convolved with a Gaussian gradient magnitude (GGM) filter  (e.g., \citealt{sanders16b}) to better show ripples and cavities. The optical Petrosian radius of the BCG's stellar component is (roughly) marked by the gray dashed ellipse. Brackets indicate the relative fields of view shown in the surrounding panels. (\textit{b}), (\textit{c}), and (\textit{d}) the same data, slightly zoomed in, with MUSE H$\alpha$, ALMA CO(2-1), and 8.4 GHz radio contours overlaid in green, blue, and gray, respectively. Moving inward, the ALMA contours in panel \textit{d} show emission
  that is $3\sigma$, $5\sigma$, $10\sigma$, and $20\sigma$ over the background
  RMS noise level. With the caveat that projection effects complicate interpretation, the H$\alpha$ nebula
  shows strong circumstantial evidence that at least some of the filaments are draped around
  the edges of the buoyant X-ray cavities marked by arrows.
 }
 \label{fig:deepchandra}
\end{figure*}

\begin{figure}
 \begin{center}
  \includegraphics[width=0.47\textwidth]{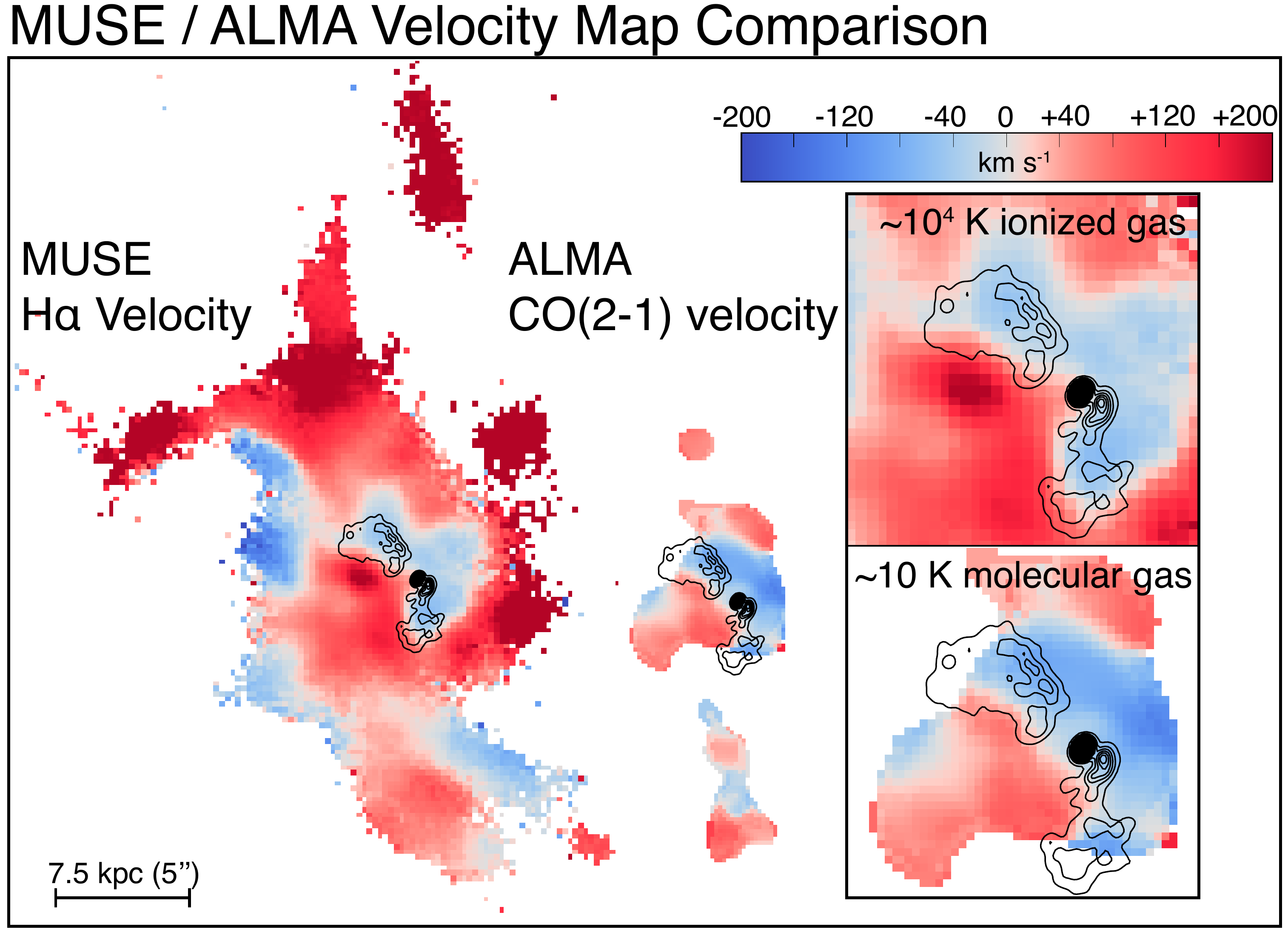}
 \end{center}
 \vspace*{-3mm}
 \caption{A side-by-side comparison of the MUSE H$\alpha$ and ALMA CO(2-1) LOS velocity maps, shown on the same spatial and velocity scales. Insets show a zoom-in on the nuclear region for both maps. Black contours again show the 8.4 GHz radio source. Where they overlap, the MUSE and ALMA velocity maps look very similar to one another. These maps are more quantitatively compared in \autoref{fig:MuseALMARatios}.  }
 \label{fig:MuseALMAVelocity}
\end{figure}

\subsection{The Fountain's ``Plume''}

There is little doubt that this pump injects an enormous amount of kinetic energy into the hot $\sim10^{7}-10^{8}$ K phase.
In \autoref{fig:deepchandra} we present a new, deep \textit{Chandra}
X-ray map of the A2597 BCG and its outskirts, made by combining the new observations
from our recent Cycle 18 Large Program with the archival exposures
previously published by \citet{mcnamara01,clarke05} and \citet{tremblay12a}.
The new map contains 1.54 million source counts collected over 626 ksec
of total integration time, enabling an exquisitely deep look
at the X-ray cavity network that permeates the innermost 30 kpc of the
cool core. The figure makes use of a Gaussian Gradient Magnitude (GGM)
filter as an edge-detector \citep{sanders16b,walker17}, revealing the X-ray cavities in sharp
relief. A discussion of detailed X-ray morphology along with
deep spectral maps, etc., will be discussed in a forthcoming paper (Tremblay et al.~in prep). We preview the map here because it makes
obvious the need to consider uplift by buoyant hot cavities
as a primary sculptor of morphology in the cold and warm nebulae.

To the north and south, H$\alpha$ filaments  (green contours on \autoref{fig:deepchandra})
appear draped over the edges of the inner X-ray cavities,
as if they have either been uplifted as they buoyantly rise,
or have formed \textit{in situ} along their wakes and rims
(e.g., \citealt{brighenti15,mcnamara16}). The nothernmost H$\alpha$ filament
has a morphology and X-ray cavity correspondence that is reminiscent
of the Northwestern ``horseshoe'' filament in Perseus (e.g., \citealt{hatch06,fabian08,gendron-marsolais18}).
In projection, the southern H$\alpha$
filaments reach a terminus at the rim of the southern cavity, forking like a snake's tongue into two thinner
filaments. As seen in the H$\alpha$ velocity map (\autoref{fig:MuseShowcase}), one filament approaches and the other recedes, yet both have a
coherent bulk line of sight velocity that is similar to the expected terminal velocity
of the buoyantly rising hot bubble with which it is cospatial (roughly half the sound speed in the hot gas, or $\sim375$ km s\mone, \citealt{tremblay12a}).
A similar ``snakes tongue'' split is seen in the redshifted northern filaments, whose $\gae15$ kpc
outskirts at the edges of cavities show the fastest LOS velocities of any optical emission line in the galaxy ($+400$ km s\mone).

The cospatial and comoving components of the warm and cold nebulae likely
trace the same population of clouds, as we have argued repeatedly throughout this paper, and as has been suggested by many authors
over many years (e.g., \citealt{odea94,jaffe97,jaffe05,wilman06,emonts13,anderson17}).
In \autoref{fig:MuseALMAVelocity} we compare the H$\alpha$ and CO(2-1) line of sight
velocity maps side-by-side.
Where they overlap, the projected velocity
of the molecular gas matches that of the warm gas, consistent
with the hypothesis that much of the Balmer emission
stems from warm ionized envelopes of cold molecular cores, tracing their interface with the ambient hot gas.
As projected on the sky, the H$\alpha$ nebula only shows line of sight velocities
consistently in excess of mean CO(2-1) velocities at galaxy-centric radii that are
greater than the outermost extent of the detected CO(2-1) emission.
Were we able to detect CO(2-1) at these large radii, we would likely find it at similar LOS velocities
as the H$\alpha$. The fact that the latter shows a factor of two broader linewidth, then, is not
necessarily surprising. Perhaps simply because of a sensitivity floor, cold molecular gas is
confined to smaller radii. Any given line of sight therefore intersects a smaller volume occupied
by CO(2-1)-bright clouds -- and therefore smaller-scale turbulent eddies -- which in turn
have smaller velocity dispersions. H$\alpha$ is both vastly brighter (i.e. easier to detect at large radii) than CO(2-1) relative
to the sensitivity limits of our optical and mm observations, respectively.
Moreover, CO(2-1)-bright molecular clouds can dissociate easily, absent sufficient shielding,
and so may be more vulnerable to destruction at larger galaxy-centric radii.
\ion{H}{1} in A2597, as mapped in detail by \citet{odea94}, shows broader linewidths more consistent
with those found in H$\alpha$, supporting this notion.

In any case, if a substantial component of the H$\alpha$ filaments have been buoyantly
uplifted in the rise of the X-ray cavities, then so too must be the molecular filaments.
Assuming (hypothetically) that coupling efficiency is not an issue,
simple energetics arguments suggest that the
cavity network in Abell 2597 is powerful enough to uplift the entirety of the cold molecular nebula.
Archimedes' principle dictates that the bubbles cannot lift more mass than they displace (e.g., \citealt{mcnamara14,russell17,vantyghem16}).
The mass of hot gas displaced in the inflation of the cavity network
is at least $\sim7\times10^9$ \Msol\ (using X-ray gas density and cavity size measurements from \citealt{tremblay12b}, assuming spherical cavity geometry, and adopting the arguments in \citealt{gitti11}), while the total cold gas mass in the molecular nebula is less than this ($\sim3.2\times10^9$ \Msol).
Moreover, the cavity system has an estimated $4pV$ mechanical energy of $\sim4\times10^{58}$ ergs \citep{tremblay12b}, while the
total kinetic energy in the cold molecular nebula (e.g., $\frac{1}{2} M_\mathrm{mol} v^2$) is about an order of magnitude lower, at roughly $\sim2\times10^{57}$ ergs. Therefore, if we ignore
coupling efficiency, uplift of the entire mass of the molecular nebula
would be safely within the kinetic energy budget of the system.

Any such uplift would be temporary. The escape speed from the galaxy,
which is roughly twice the circular speed at any given radius, is far in excess
of any observed line of sight velocity in the system. After decoupling from either
the cavity wake or jet entrainment layer that has lifted them to higher altitudes, cold clouds
should fall back inward at their terminal speed, drifting in the hot gas velocity
field as they descend. These infalling clouds may join the population we observe
in absorption, powering black hole activity once again, and keeping the
fountain long-lived.

%
%
%

\acknowledgments This paper makes use of the following ALMA data:
ADS/JAO.ALMA\#2012.1.00988.S. ALMA is a partnership of ESO (representing its member states),
NSF (USA) and NINS (Japan), together with NRC (Canada) and NSC and ASIAA (Taiwan),
in cooperation with the Republic of Chile. The Joint ALMA Observatory
is operated by ESO, AUI/NRAO and NAOJ. We are grateful to the European ALMA Regional Centres,
particularly
those in Garching and Manchester,
for their dedicated end-to-end support of data associated with this paper.
We have also received immense support from the National Radio Astronomy Observatory,
a facility of the National Science Foundation operated under cooperative agreement by Associated Universities, Inc.

This work is also based on observations made with ESO Telescopes at the La Silla Paranal Observatory under programme ID 094.A-0959 (PI: Hamer).
We also present observations made with the NASA/ESA \textit{Hubble Space Telescope},
obtained from the data archive at the Space Telescope Science Institute (STScI).
STScI is operated by the Association of Universities for Research in Astronomy, Inc. under NASA contract NAS 5-26555.

GRT thanks EST and AST for educating him on Nature's many sources of uplift.
GRT also acknowledges support from the National Aeronautics
and Space Administration (NASA) through
\textit{Chandra} Award Number GO7-8128X as well as Einstein Postdoctoral Fellowship
Award Number PF-150128, issued by the Chandra X-ray Center, which is operated by the Smithsonian Astrophysical
Observatory for and on behalf of NASA under contract NAS8-03060.
BJW, SWR, JAZ PEJN, RPK, WRF, CJ, and YS also acknowledge the financial support of NASA contract NAS8-03060 (Chandra X-ray Center).
FC acknowledges the European Research Council for the Advanced Grant Program \# 267399-Momentum.
MG is supported by NASA through Einstein Postdoctoral Fellowship Award Number PF5-160137, as well as \textit{Chandra} grant GO7-18121X.
The work of SAB, CPO, and BRM was supported by a generous grant from the Natural Sciences and Engineering Research Council of Canada.
HRR and TAD acknowledge support from a Science and Technology Facilities Council (STFC) Ernest Rutherford Fellowship.
ACE acknowledges support from STFC grant ST/P00541/1.
ACF acknowledges support from ERC Advanced Grant `Feedback'.
MNB acknowledges funding from the STFC. Basic research in radio astronomy at the Naval Research Laboratory is supported by 6.1 Base funding.
This research made use of \texttt{Astropy}\footnote{\url{http://www.astropy.org/}}, a community-developed core Python package for Astronomy \citep{astropypaper, astropypaper2}.
Some MUSE data reduction and analysis was conducted on \textit{Hydra}, the Smithsonian Institution's High Performance Cluster (SI/HPC).

\facilities{
 CXO (ACIS-S),
 HST (ACS, NICMOS),
 VLT: Yepun
}

\software{%
 \texttt{Astropy} \citep{astropypaper, astropypaper2},
 \texttt{CASA} \citep{mcmullin07},
 \texttt{CIAO} \citep{fruscione06},
 \texttt{IPython} \citep{ipython},
 \texttt{Matplotlib} \citep{matplotlib},
 \texttt{NumPy} \citep{numpy},
 \texttt{PySpecKit} \citep{pyspeckit},
 \texttt{scipy} \citep{scipy}
}

\dataset[DOI-linked Software Repository for this Paper]{http://doi.org/10.5281/zenodo.1233825}

\bibliography{grant_refs}

\end{document}